\documentclass[aps,prd,amsmath,amssymb,reprint,showpacs,superscriptaddress,nobibnotes,floatfix]{revtex4-1}

\usepackage{multirow}
\usepackage{esvect}
\usepackage[tracking=true,kerning=true,expansion=true,spacing=false,factor=1100,stretch=20,shrink=20]{microtype}
\SetTracking{encoding={*}, shape=sc}{50}

\usepackage{siunitx}
\DeclareSIUnit\clight{\text{\ensuremath{c}}}

\usepackage[pdftex,unicode=true,pdfpagelabels=true,pdfusetitle,pdfauthor={The CDF Collaboration},hidelinks]{hyperref}

\usepackage{tikz}

\newcommand{\columnwidthfigure}{0.90\columnwidth}

\newcommand{\afbtt}{A_{\text{FB}}^{t\bar{t}}}
\newcommand{\dil}{\ttbar \rightarrow { \ell^{+}\ell^{-}}\nu\bar{\nu} b\bar{b}}
\newcommand{\lj}{\ttbar \rightarrow \ell\nu q\bar{q}b\bar{b}}
\newcommand{\allhad}{\ttbar \rightarrow qq\bar{q}\bar{q}b\bar{b}}

\newcommand{\Me}{\mbox{$E\kern-0.50em\raise0.10ex\hbox{/}$}}
\newcommand{\pztt}{p_{z, \ttbar}}
\newcommand{\pttt}{p_{T, \ttbar}}
\newcommand{\Mtt}{m_{\ttbar}}

\newcommand{\MetVec}{\mbox{$\vec E\kern-0.50em\raise0.10ex\hbox{/}_{T}$}}
\newcommand{\MetVecRaw}{\mbox{$\vec E\kern-0.50em\raise0.10ex\hbox{/}^{raw}_{T}$}}
\newcommand{\Met}{\mbox{$E\kern-0.50em\raise0.10ex\hbox{/}_{T}$}}
\newcommand{\Mex}{\mbox{$E\kern-0.50em\raise0.10ex\hbox{/}_{x}$}}
\newcommand{\Mey}{\mbox{$E\kern-0.50em\raise0.10ex\hbox{/}_{y}$}}
\newcommand{\Mexy}{\mbox{$E\kern-0.50em\raise0.10ex\hbox{/}_{x,y}$}}
\newcommand{\MetRaw}{\mbox{$E\kern-0.50em\raise0.10ex\hbox{/}_{T}^{raw}$}}
\newcommand{\MetSpec}{\mbox{$E\kern-0.50em\raise0.10ex\hbox{/}_{T}^{spec}$}}

\newcommand{\MetSig}{\mbox{$E\kern-0.50em\raise0.10ex\hbox{/}_{T}^{sig}$}}
\newcommand{\fbm}{\rm{fb}^{-1}}

\newcommand{\ttbar}{t \bar{t}}

\newcommand{\gev}{\mathrm{GeV}}

\newcommand{\gevcc}{\mathrm{GeV}/c^2}

\newcommand{\pt}{p_{T}}

\newcommand{\MET}{\mbox{$E\kern-0.50em\raise0.10ex\hbox{/}_{T}$}}
\newcommand{\met}{\mbox{$E\kern-0.50em\raise0.10ex\hbox{/}_{T}$}}
\newcommand{\METzero}{\mbox{$E\kern-0.50em\raise0.10ex\hbox{/}_{T}^0$}}
\newcommand{\vecMET}{\mbox{$\vec{E}\kern-0.50em\raise0.10ex\hbox{/}_{T}$}}
\newcommand{\METSPEC}{\mbox{$E\kern-0.50em\raise0.10ex\hbox{/}_{Tspec}$}}

\newcommand{\ppbar}{p \overline{p}}

\def\slantfrac#1#2{\kern.1em^{#1}\kern-.3em/\kern-.1em_{#2}}
\newcommand{\syst}{\text{syst}}
\newcommand{\stat}{\text{stat}}
\newcommand{\et}{E_{{T}}}

\newcommand{\afblep}{A_{\text{FB}}^{\ell}}
\newcommand{\afbdeta}{A_{\text{FB}}^{\ell\ell}}
\newcommand{\dy}{\Delta y}

\newcommand{\jd}{\delta_{\text{j}}}
\newcommand{\jdpeak}{\delta_{\text{j,peak}}}
\newcommand{\nparton}{N_{\text{parton}}}
\newcommand{\nexp}{N_{\text{exp}}}
\newcommand{\eff}{\epsilon}
\newcommand{\detS}{S}
\newcommand{\nbkg}{N_{\text{bkg}}}
\newcommand{\nobs}{N_{\text{obs}}}

\mathchardef\mhyphen="2D

\begin{document}
\newcommand{\TableCutWeightOptimal}{
\begin{ruledtabular}
\begin{tabular}{lc}
    Quantity&Optimal value\\
    \hline
    $\Theta(\jd)$ & 3.5\\
    $w_{Q}$ & 0.7\\
    $\Theta(m_{lb}^{2})$ & 24000 $\gev^{2}/c^{4}$\\
    $\Theta(\Delta R_{\text{min}})$ & 0.2\\
\end{tabular}
\end{ruledtabular}
}

\newcommand\Tstrut{\rule{0pt}{2.4ex}}       
\newcommand\Bstrut{\rule[-1.3ex]{0pt}{0pt}} 

\newcommand{\TableEventYield}{
\begin{tabular*}{\columnwidth}{l@{\extracolsep{\fill}}
     S[table-format=1.1(1),table-align-uncertainty=true]}
\hline\hline
Source & {Events}\\\hline
 Diboson & 26+-5\\
 $Z/\gamma^{*}$+jets&37+-4\\
 $W$+jets&28+-9\\
 $\ttbar$ non-dilepton&5.3+-0.3\\
 Total background&96+-18\\
 Signal $\ttbar~(\sigma = 7.4~\mathrm{pb})$&386+-18\\
 Total SM expectation & 482+-36\\
 Observed&{495}\\
\hline\hline
\end{tabular*}
}

\newcommand{\TableUncertaintyDifferential}{
\begin{ruledtabular}
\begin{tabular}{lccc}
Source of uncertainty&$\afbtt(\text{inclusive})$&$\afbtt (|\dy|<0.5)$&$\afbtt (|\dy|>0.5)$\\\hline
Statistical uncertainty&0.11&0.33&0.13\\\hline
Background&0.04&0.13&0.06\\
Parton showering&0.03&0.07&0.06\\
Color reconnection&0.03&0.12&0.06\\
Initial- and final-state radiation&0.03&0.05&0.03\\
Jet energy scale&0.02&0.02&0.02\\
NLO assumption&0.02& 0.06&0.02\\
Parton-distribution functions&0.01& 0.01&0.01\\\hline
Total systematic uncertainty&0.07& 0.20&0.11\\\hline
Total uncertainty&0.13&0.39&0.17\\
\end{tabular}
\end{ruledtabular}
}

\newcommand{\TableUncertaintyCorrelation}{
\begin{ruledtabular}
\begin{tabular}{lccc}
Source of uncertainty&Lepton+jets &Dilepton &Correlation\\\hline
Background shape & 0.018 & \multirow{2}{*}{0.04}&\multirow{2}{*}{0}\\
Background normalization&0.013&&\\
Parton shower&0.01&0.03&1\\
Jet energy scale&0.007&0.02&1\\
Inital- and final-state radiation&0.005&0.03&1\\
Correction procedure / NLO assumption &0.004&0.02&0\\
Color reconnection&0.001&0.03&1\\
Partion-distribution functions&0.001&0.01&1\\\hline
Total systematic uncertainty&0.026&0.07&\\
Statistical uncertainty& 0.039 & 0.11&0\\\hline
Total uncertainty&0.047&0.13&\\
\end{tabular}
\end{ruledtabular}
}

\newcommand{\TableDifferentialCombinationBinCentroidAFB}{
\begin{ruledtabular}
\begin{tabular}{lcccccc}
\multirow{2}{*}{}& \multicolumn{4}{c}{Lepton+jets}&\multicolumn{2}{c}{Dilepton}\\\cline{2-5}\cline{6-7}
& $|\dy|<0.5$ & $0.5<|\dy|<1.0$& $1.0<|\dy|<1.5$& $|\dy|>1.5$& $|\dy|<0.5$& $|\dy|>0.5$\\\hline
Bin centroid & 0.24&0.73&1.22&1.82&0.24&1.01\\
$\afbtt(|\dy|)$ & 0.048&0.180&0.356&0.477&0.11&0.13 
\end{tabular}
\end{ruledtabular}
}

\newcommand{\TableDifferentialCombinationCovMat}{
\begin{ruledtabular}
\begin{tabular}{lcccccc}
& \multicolumn{4}{c}{Lepton+jets}&\multicolumn{2}{c}{Dilepton}\\\cline{2-5}\cline{6-7}
\multirow{1}{*}{Eigenvalue $\lambda$}& $|\dy|<0.5$ & $0.5<|\dy|<1.0$& $1.0<|\dy|<1.5$& $|\dy|>1.5$& $|\dy|<0.5$& $|\dy|>0.5$\\\hline
0.156    & -0.018&0.001&0.008&0.030&-0.984&0.174\\
0.0296   & 0.064&-0.030&-0.440&-0.830&-0.087&-0.322\\
0.0251   & -0.012&-0.014&-0.172&-0.286&0.155&0.930\\
0.00732  & -0.371&-0.840&-0.344&0.193&0.005&-0.023\\
0.000682 & 0.904&-0.235&-0.281&0.219&-0.008&0.024\\
0.000476 & -0.201&0.487&-0.761&0.378&0.006&-0.021\\
\end{tabular}
\end{ruledtabular}
}

\affiliation{Institute of Physics, Academia Sinica, Taipei, Taiwan 11529, Republic of China}
\affiliation{Argonne National Laboratory, Argonne, Illinois 60439, USA}
\affiliation{University of Athens, 157 71 Athens, Greece}
\affiliation{Institut de Fisica d'Altes Energies, ICREA, Universitat Autonoma de Barcelona, E-08193, Bellaterra (Barcelona), Spain}
\affiliation{Baylor University, Waco, Texas 76798, USA}
\affiliation{Istituto Nazionale di Fisica Nucleare Bologna, \ensuremath{^{kk}}University of Bologna, I-40127 Bologna, Italy}
\affiliation{University of California, Davis, Davis, California 95616, USA}
\affiliation{University of California, Los Angeles, Los Angeles, California 90024, USA}
\affiliation{Instituto de Fisica de Cantabria, CSIC-University of Cantabria, 39005 Santander, Spain}
\affiliation{Carnegie Mellon University, Pittsburgh, Pennsylvania 15213, USA}
\affiliation{Enrico Fermi Institute, University of Chicago, Chicago, Illinois 60637, USA}
\affiliation{Comenius University, 842 48 Bratislava, Slovakia; Institute of Experimental Physics, 040 01 Kosice, Slovakia}
\affiliation{Joint Institute for Nuclear Research, RU-141980 Dubna, Russia}
\affiliation{Duke University, Durham, North Carolina 27708, USA}
\affiliation{Fermi National Accelerator Laboratory, Batavia, Illinois 60510, USA}
\affiliation{University of Florida, Gainesville, Florida 32611, USA}
\affiliation{Laboratori Nazionali di Frascati, Istituto Nazionale di Fisica Nucleare, I-00044 Frascati, Italy}
\affiliation{University of Geneva, CH-1211 Geneva 4, Switzerland}
\affiliation{Glasgow University, Glasgow G12 8QQ, United Kingdom}
\affiliation{Harvard University, Cambridge, Massachusetts 02138, USA}
\affiliation{Division of High Energy Physics, Department of Physics, University of Helsinki, FIN-00014, Helsinki, Finland; Helsinki Institute of Physics, FIN-00014, Helsinki, Finland}
\affiliation{University of Illinois, Urbana, Illinois 61801, USA}
\affiliation{The Johns Hopkins University, Baltimore, Maryland 21218, USA}
\affiliation{Institut f\"{u}r Experimentelle Kernphysik, Karlsruhe Institute of Technology, D-76131 Karlsruhe, Germany}
\affiliation{Center for High Energy Physics: Kyungpook National University, Daegu 702-701, Korea; Seoul National University, Seoul 151-742, Korea; Sungkyunkwan University, Suwon 440-746, Korea; Korea Institute of Science and Technology Information, Daejeon 305-806, Korea; Chonnam National University, Gwangju 500-757, Korea; Chonbuk National University, Jeonju 561-756, Korea; Ewha Womans University, Seoul, 120-750, Korea}
\affiliation{Ernest Orlando Lawrence Berkeley National Laboratory, Berkeley, California 94720, USA}
\affiliation{University of Liverpool, Liverpool L69 7ZE, United Kingdom}
\affiliation{University College London, London WC1E 6BT, United Kingdom}
\affiliation{Centro de Investigaciones Energeticas Medioambientales y Tecnologicas, E-28040 Madrid, Spain}
\affiliation{Massachusetts Institute of Technology, Cambridge, Massachusetts 02139, USA}
\affiliation{University of Michigan, Ann Arbor, Michigan 48109, USA}
\affiliation{Michigan State University, East Lansing, Michigan 48824, USA}
\affiliation{Institution for Theoretical and Experimental Physics, ITEP, Moscow 117259, Russia}
\affiliation{University of New Mexico, Albuquerque, New Mexico 87131, USA}
\affiliation{The Ohio State University, Columbus, Ohio 43210, USA}
\affiliation{Okayama University, Okayama 700-8530, Japan}
\affiliation{Osaka City University, Osaka 558-8585, Japan}
\affiliation{University of Oxford, Oxford OX1 3RH, United Kingdom}
\affiliation{Istituto Nazionale di Fisica Nucleare, Sezione di Padova, \ensuremath{^{ll}}University of Padova, I-35131 Padova, Italy}
\affiliation{University of Pennsylvania, Philadelphia, Pennsylvania 19104, USA}
\affiliation{Istituto Nazionale di Fisica Nucleare Pisa, \ensuremath{^{mm}}University of Pisa, \ensuremath{^{nn}}University of Siena, \ensuremath{^{oo}}Scuola Normale Superiore, I-56127 Pisa, Italy, \ensuremath{^{pp}}INFN Pavia, I-27100 Pavia, Italy, \ensuremath{^{qq}}University of Pavia, I-27100 Pavia, Italy}
\affiliation{University of Pittsburgh, Pittsburgh, Pennsylvania 15260, USA}
\affiliation{Purdue University, West Lafayette, Indiana 47907, USA}
\affiliation{University of Rochester, Rochester, New York 14627, USA}
\affiliation{The Rockefeller University, New York, New York 10065, USA}
\affiliation{Istituto Nazionale di Fisica Nucleare, Sezione di Roma 1, \ensuremath{^{rr}}Sapienza Universit\`{a} di Roma, I-00185 Roma, Italy}
\affiliation{Mitchell Institute for Fundamental Physics and Astronomy, Texas A\&M University, College Station, Texas 77843, USA}
\affiliation{Istituto Nazionale di Fisica Nucleare Trieste, \ensuremath{^{ss}}Gruppo Collegato di Udine, \ensuremath{^{tt}}University of Udine, I-33100 Udine, Italy, \ensuremath{^{uu}}University of Trieste, I-34127 Trieste, Italy}
\affiliation{University of Tsukuba, Tsukuba, Ibaraki 305, Japan}
\affiliation{Tufts University, Medford, Massachusetts 02155, USA}
\affiliation{Waseda University, Tokyo 169, Japan}
\affiliation{Wayne State University, Detroit, Michigan 48201, USA}
\affiliation{University of Wisconsin-Madison, Madison, Wisconsin 53706, USA}
\affiliation{Yale University, New Haven, Connecticut 06520, USA}

\author{T.~Aaltonen}
\affiliation{Division of High Energy Physics, Department of Physics, University of Helsinki, FIN-00014, Helsinki, Finland; Helsinki Institute of Physics, FIN-00014, Helsinki, Finland}
\author{S.~Amerio\ensuremath{^{ll}}}
\affiliation{Istituto Nazionale di Fisica Nucleare, Sezione di Padova, \ensuremath{^{ll}}University of Padova, I-35131 Padova, Italy}
\author{D.~Amidei}
\affiliation{University of Michigan, Ann Arbor, Michigan 48109, USA}
\author{A.~Anastassov\ensuremath{^{w}}}
\affiliation{Fermi National Accelerator Laboratory, Batavia, Illinois 60510, USA}
\author{A.~Annovi}
\affiliation{Laboratori Nazionali di Frascati, Istituto Nazionale di Fisica Nucleare, I-00044 Frascati, Italy}
\author{J.~Antos}
\affiliation{Comenius University, 842 48 Bratislava, Slovakia; Institute of Experimental Physics, 040 01 Kosice, Slovakia}
\author{G.~Apollinari}
\affiliation{Fermi National Accelerator Laboratory, Batavia, Illinois 60510, USA}
\author{J.A.~Appel}
\affiliation{Fermi National Accelerator Laboratory, Batavia, Illinois 60510, USA}
\author{T.~Arisawa}
\affiliation{Waseda University, Tokyo 169, Japan}
\author{A.~Artikov}
\affiliation{Joint Institute for Nuclear Research, RU-141980 Dubna, Russia}
% ...

\collaboration{CDF Collaboration}
\altaffiliation[With visitors from]{
\ensuremath{^{a}}University of British Columbia, Vancouver, BC V6T 1Z1, Canada,
\ensuremath{^{b}}Istituto Nazionale di Fisica Nucleare, Sezione di Cagliari, 09042 Monserrato (Cagliari), Italy,
\ensuremath{^{c}}University of California Irvine, Irvine, CA 92697, USA,
\ensuremath{^{d}}Institute of Physics, Academy of Sciences of the Czech Republic, 182~21, Czech Republic,
\ensuremath{^{e}}CERN, CH-1211 Geneva, Switzerland,
\ensuremath{^{f}}Cornell University, Ithaca, NY 14853, USA,
\ensuremath{^{g}}University of Cyprus, Nicosia CY-1678, Cyprus,
\ensuremath{^{h}}Office of Science, U.S. Department of Energy, Washington, DC 20585, USA,
\ensuremath{^{i}}University College Dublin, Dublin 4, Ireland,
\ensuremath{^{j}}ETH, 8092 Z\"{u}rich, Switzerland,
\ensuremath{^{k}}University of Fukui, Fukui City, Fukui Prefecture, Japan 910-0017,
\ensuremath{^{l}}Universidad Iberoamericana, Lomas de Santa Fe, M\'{e}xico, C.P. 01219, Distrito Federal,
\ensuremath{^{m}}University of Iowa, Iowa City, IA 52242, USA,
\ensuremath{^{n}}Kinki University, Higashi-Osaka City, Japan 577-8502,
\ensuremath{^{o}}Kansas State University, Manhattan, KS 66506, USA,
\ensuremath{^{p}}Brookhaven National Laboratory, Upton, NY 11973, USA,
\ensuremath{^{q}}Istituto Nazionale di Fisica Nucleare, Sezione di Lecce, Via Arnesano, I-73100 Lecce, Italy,
\ensuremath{^{r}}Queen Mary, University of London, London, E1 4NS, United Kingdom,
\ensuremath{^{s}}University of Melbourne, Victoria 3010, Australia,
\ensuremath{^{t}}Muons, Inc., Batavia, IL 60510, USA,
\ensuremath{^{u}}Nagasaki Institute of Applied Science, Nagasaki 851-0193, Japan,
\ensuremath{^{v}}National Research Nuclear University, Moscow 115409, Russia,
\ensuremath{^{w}}Northwestern University, Evanston, IL 60208, USA,
\ensuremath{^{x}}University of Notre Dame, Notre Dame, IN 46556, USA,
\ensuremath{^{y}}Universidad de Oviedo, E-33007 Oviedo, Spain,
\ensuremath{^{z}}CNRS-IN2P3, Paris, F-75205 France,
\ensuremath{^{aa}}Universidad Tecnica Federico Santa Maria, 110v Valparaiso, Chile,
\ensuremath{^{bb}}Sejong University, Seoul 143-747, Korea,
\ensuremath{^{cc}}The University of Jordan, Amman 11942, Jordan,
\ensuremath{^{dd}}Universite catholique de Louvain, 1348 Louvain-La-Neuve, Belgium,
\ensuremath{^{ee}}University of Z\"{u}rich, 8006 Z\"{u}rich, Switzerland,
\ensuremath{^{ff}}Massachusetts General Hospital, Boston, MA 02114 USA,
\ensuremath{^{gg}}Harvard Medical School, Boston, MA 02114 USA,
\ensuremath{^{hh}}Hampton University, Hampton, VA 23668, USA,
\ensuremath{^{ii}}Los Alamos National Laboratory, Los Alamos, NM 87544, USA,
\ensuremath{^{jj}}Universit\`{a} degli Studi di Napoli Federico II, I-80138 Napoli, Italy
}
\noaffiliation

\title{ Measurement of the forward--backward asymmetry of top-quark and antiquark pairs using the full CDF Run II data set}

\begin{abstract}
    We measure the forward--backward asymmetry of the production of top quark and antiquark pairs in proton-antiproton collisions at center-of-mass energy $\sqrt{s} = 1.96~\mathrm{TeV}$ using the full data set collected by the Collider Detector at Fermilab (CDF) in Tevatron Run II corresponding to an integrated luminosity of $9.1~\fbm$. The asymmetry is characterized by the rapidity difference between top quarks and antiquarks ($\dy$), and measured in the final state with two charged leptons (electrons and muons).
    The inclusive asymmetry, corrected to the entire phase space at parton level, is measured to be $\afbtt = 0.12 \pm 0.13$, consistent with the expectations from the standard-model (SM) and previous CDF results in the final state with a single charged lepton.  
    The combination of the CDF measurements of the inclusive $\afbtt$ in both final states yields $\afbtt=0.160\pm0.045$, which is consistent with the SM predictions. 
    We also measure the differential asymmetry as a function of $\dy$.
    A linear fit to $\afbtt(|\dy|)$, assuming zero asymmetry at $\dy=0$, yields a slope of $\alpha=0.14\pm0.15$, consistent with the SM prediction 
    and the previous CDF determination in the final state with a single charged lepton.
    The combined slope of $\afbtt(|\dy|)$ in the two final states is $\alpha=0.227\pm0.057$, which is $2.0\sigma$ larger than the SM prediction.
\end{abstract}

\date{\today}

\maketitle
\clearpage

\newpage

\section{\label{sec:Intro}Introduction}

The forward--backward asymmetry of the production of top-quark and antiquark pairs ($\ttbar$) in high-energy proton-antiproton collisions is an observable unique to the Tevatron experiments. It quantifies the preference of top quarks to follow the proton direction, ``forward," instead of the antiproton direction, ``backward."
At leading order (LO), quantum chromodynamics (QCD) predicts no net asymmetry in $\ttbar$ production. All asymmetric effects come from interference effects with electroweak and higher-order QCD amplitudes~\cite{Bernreuther:2012sx}. The top-quark-pair forward--backward asymmetry ($\afbtt$) measurement program at the Tevatron uses the proton-antiproton initial state with center-of-mass energy at $1.96~\mathrm{TeV}$ to probe both the higher-order effects of the standard model (SM) and scenarios beyond the SM. This complements the precision measurements of top-quark physics at the LHC, where top-quark-pair production is dominated by gluon-gluon interactions and is therefore symmetric along the beamline direction~\cite{Bernreuther:2012sx}. There were tensions between the experimental results of $\afbtt$~\cite{cdf_afb_prd2011,d0_afb_prd2011} and the SM theoretical calculations~\cite{Kuhn:1998kw,PhysRevD.84.093003,Manohar2012313,jhep012012063,Bernreuther:2012sx}.
This article reports the final measurement of $\afbtt$ by the Collider Detector at Fermilab (CDF) experiment.

We define $\afbtt$ as
\begin{equation}
    \afbtt = \frac{N(\dy > 0) - N(\dy < 0)}{N(\dy > 0) + N(\dy < 0)}, \label{eq:afbtt}
\end{equation}
where $N$ is the number of $\ttbar$ pairs, $y$ is the rapidity of the top quark ($y_{t}$) or antiquark ($y_{\bar{t}}$) defined relative to the proton beam direction~\cite{geometry}, and $\dy = y_{t}-y_{\bar{t}}$.
A next-to-next-to-leading order (NNLO) calculation yields the prediction $\afbtt=0.095\pm0.007$~\cite{Czakon:2014xsa}, which becomes $\afbtt=0.100\pm0.006$ after adding soft-gluon corrections~\cite{PhysRevD.91.071502}. The predicted asymmetry is greatly enhanced in certain kinematic regions, like the high $\ttbar$ invariant-mass region or the high-$|\dy|$ region, thus measurements of differential asymmetries are also of great importance~\cite{Czakon:2016ckf}. 
If non-SM particles contribute to the dynamics, the asymmetry could be significantly changed~\cite{Aguilar-Saavedra:2014kpa}.

Measurements of the inclusive $\afbtt$, corrected to the entire phase space at parton level, can be made using top quark-antiquark pairs that yield final states with either a charged lepton ($\ell$) and four hadronic jets from collimated cluster of incident hadrons from light ($q$) and bottom ($b$) quarks ($\lj$, or lepton+jets) or two charged leptons and two bottom-quark jets ($\dil$, or dilepton). The $\afbtt$ measurement in the $\allhad$ (all-hadronic) final state is not practical at the Tevatron experiments due to the experimental difficulties in determining the charge of the quarks that initiate the jets~\cite{PhysRevD.88.032003}. With the CDF data, corresponding to $9.4~\fbm$ of integrated luminosity, the measurement in the lepton+jets final state yielded a value of $0.164\pm0.047$~\cite{Aaltonen:2012it}, which is consistent with the NNLO SM prediction~\cite{Czakon:2014xsa} within $1.5\sigma$. The same measurements with data from the D0 collaboration corresponding to $9.7~\fbm$ of integrated luminosity in the lepton+jets~\cite{Abazov:2014cca} and dilepton final state~\cite{Abazov:2015fna} yielded $0.106\pm0.030$ and $0.175\pm0.063$, respectively, which are consistent with the NNLO SM prediction~\cite{Czakon:2014xsa}.
The differential $\afbtt$ measurements as functions of the invariant mass of $\ttbar$ ($\Mtt$) and $\dy$ at CDF in the lepton+jets final state~\cite{Aaltonen:2012it} showed mild tension with the SM predictions, while the results in the D0 lepton+jets final state~\cite{Abazov:2014cca} showed consistency. 

The leptons from the top-quark cascade decays carry directional information from their parent top quarks, and thus forward--backward asymmetry measurements of the leptons ($\afblep$ and $\afbdeta$) serve as complementary measurements to $\afbtt$~\cite{Bernreuther201090}. Results from the CDF dilepton final state and the D0 lepton+jets and dilepton final states mostly showed agreement with the SM, whereas the CDF lepton+jets result showed mild tension with the SM~\cite{Aaltonen:2013vaf,Aaltonen:2014eva,Abazov:2014oea,Abazov:2013wxa}.

Additionally, a more detailed study of the cross section of the $\ttbar$ system as a function of the production angle of the top quark relative to the proton direction in the $\ttbar$ rest frame ($\theta^{*}$) was performed in the lepton+jets final state at CDF~\cite{PhysRevLett.111.182002}. The differential cross section $d\sigma/d\cos\theta^{*}$ was expanded in Legendre polynomials and the mild asymmetry enhancement was attributed to the term linear in $\cos\theta^{*}$.
Since many features of top-quark-pair production are well described by the SM, such as the inclusive cross section~\cite{PhysRevD.89.072001} and the differential cross sections as functions of the transverse momentum of the top quarks ($p_{T,t}$), $\Mtt$, etc.~\cite{PhysRevD.90.092006}, any contribution from non-SM dynamics that would affect the top-quark asymmetry would need to have minimal impact on these properties to preserve consistency with experimental constraints. Therefore, we use an \textit{ad hoc} procedure suggested by the $d\sigma/d\cos\theta^{*}$ measurement to generically explore the variations in $\afbtt$ that are consistent with all other experimental constraints. 

This article describes the measurements of the inclusive and differential $\afbtt$ values in the dilepton final state as well as their combination with the lepton+jets results. These measurements use the entire data set collected by the CDF detector during Tevatron Run II, corresponding to an integrated luminosity of $9.1~\fbm$. 
The chief experimental challenges are: 1) the reconstruction of the signal kinematic properties needed to calculate the observed asymmetry, and 2) the transformation of the observed asymmetry, derived from experimentally-observed quantities, into the parton-level asymmetry, which requires corrections for experimental effects. 
The reconstruction of the kinematic properties is complicated by the presence of two final-state neutrinos and the ambiguity in associating the $b$ and $\bar{b}$ jet with the lepton of appropriate charge. 
The two final-state neutrinos leave the kinematic properties of the signal experimentally underconstrained, introducing assumptions and reconstruction ambiguities that degrade the precision of the measurement. For each event, we use observed kinematic quantities and probability densities derived from simulation to construct a kinematic likelihood that is a function of the unobservable quantities. From that we extract the probability-density distribution for the top-quark rapidity difference. 
In addition, the difficulty in determining, event-by-event, whether a $b$ jet originates from a bottom quark or antiquark introduces a further two-fold ambiguity in the proper reconstruction of the $W$-boson decays, due to the two possible lepton-jet pairings. 
We reduce the degradation of the results due to these reconstruction difficulties by means of an optimization. This aims at minimizing the total uncertainty as evaluated by repeating the analysis on several ensembles of simulated experiments that mimic the actual experimental conditions. As a result, improved selection criteria and use of both lepton-jet pairings for each event, opportunely weighted, leads to an 11\% expected improvement in the total uncertainty.
Finally, determination of parton-level results from the observed asymmetries is achieved with a Bayesian-inference technique tested and tuned using simulated samples under various configurations.

The outline of the article is as follows: The dilepton analysis uses the same data set as Ref.~\cite{Aaltonen:2014eva}, and is summarized in Sec.~\ref{sec:TOPDILEventSelection}. A series of scenarios with various $\afbtt$ values, including those inspired by the $d\sigma/d\cos\theta^{*}$ measurement, is also described in this section.
The top-quark-pair reconstruction of the momenta of the top (anti-)quarks is described in Sec.~\ref{sec:TopReco}. We estimate the parton-level results using the Bayesian-inference technique described in Sec.~\ref{sec:Unfolding}, and employ an optimization procedure to minimize the expected uncertainties on the inclusive measurement of $\afbtt$ as illustrated in Sec.~\ref{sec:optimization}. Validations of the reconstruction and correction methodology are shown in Sec.~\ref{sec:validation}. The estimation of systematic uncertainties is described in Sec.~\ref{sec:Systematics}. We present the final measurements of both the inclusive $\afbtt$ and the differential measurement of $\afbtt$ as a function of $|\dy|$ in Sec.~\ref{sec:results}. The combination of the dilepton results and the lepton+jets results is shown in Sec.~\ref{sec:CDFComb}, followed by conclusions in Sec.~\ref{sec:conclusion}.

\section{\label{sec:TOPDILEventSelection}Detector Description, Event selection, and Signal and Background Estimation}
The CDF II detector is a general purpose, azimuthally and forward-backward symmetric magnetic spectrometer
with calorimeters and muon detectors~\cite{PhysRevD.71.032001}. Charged-particle trajectories (tracks) are reconstructed with a silicon microstrip detector and a large open-cell drift chamber in a $1.4~\mathrm{T}$ solenoidal magnetic field. Projective-tower-geometry electromagnetic and hadronic calorimeters located
outside the solenoid provide electron, jet, and missing transverse energy ($\met$) detection~\cite{MET}. Electrons are identified by matching isolated tracks to clusters of energy deposited in the electromagnetic calorimeter. Jets are identified as narrow clusters of energy deposits in the calorimeters consistent with collimated
clusters of incident hadrons. A non-zero missing transverse energy indicates an imbalance in the total event transverse momentum~\cite{MET}. Beyond the calorimeters are multilayer proportional chambers that provide muon detection and identification in the psuedorapidity~\cite{geometry} region $|\eta|\leq1.0$.  

The standard event-selection criteria for top-quark measurements in the dilepton final state at CDF are used following Ref.~\cite{Aaltonen:2014eva}. We require two oppositely-charged leptons (electrons and muons), two or more jets, and a large amount of $\met$. Other kinematic requirements are made to enhance the signal purity, to ensure good measurement of the event properties, and to ensure the robust estimate of the backgrounds~\cite{Aaltonen:2014eva}. We refer to these requirements as the baseline event-selection criteria. We add more requirements, described in Sec.~\ref{sec:optimization}, to further improve the measurement sensitivity based on the quality of top-quark-pair reconstruction.

Signal and background estimations also follow Ref.~\cite{Aaltonen:2014eva}. The $\ppbar\rightarrow \dil$ signal is modeled with the NLO Monte Carlo (MC) generator \textsc{powheg}~\cite{Alioli:2011as,*Nason:2004rx,
*Frixione:2007vw,*Alioli:2010xd}, with parton hadronization modeled by \textsc{pythia}~\cite{Sjostrand:2006za}, and a detailed simulation of the response of the CDF II detector~\cite{Gerchtein:2003ba}. Background sources include the production of
a $Z$ boson or virtual photon in association with jets ($Z/\gamma^{*} + \mathrm{jets}$), production of a $W$ boson in association with jets ($W+\mathrm{jets}$), diboson production ({\it WW}, {\it WZ}, {\it ZZ}, and $W\gamma$), and $\ttbar$ production where one of the $W$ bosons from the top-quark pair decays hadronically and one jet is misidentified as a lepton ($\ttbar$ non-dilepton). Most sources of background are modeled using simulation with the same CDF II detector simulation as used for the signal~\cite{Gerchtein:2003ba}, while the $W+\mathrm{jets}$ background is modeled using data~\cite{Aaltonen:2013tsa}. With these estimations of signal and backgrounds, we expect the baseline data set to be $568\pm40$ events, with  72\% of the contribution from signal, and we observe 569 events in the baseline data set.

In this analysis we use two categories of MC samples with various assumed $\afbtt$ values to develop and validate the measurement procedure. The first contains ensembles of simulated samples, each generated with a different choice for the true $\afbtt$, relying on the measurement of top-quark differential cross section~\cite{PhysRevLett.111.182002}. This measurement suggested that the potential $\afbtt$ excess could be due to an additional contribution to the linear term of $d\sigma/d\cos\theta^{*}$. Samples with genuine asymmetries in the range $-0.1<\afbtt<0.3$ are simulated by reweighting the \textsc{powheg} sample with appropriate additional linear contributions as functions of $\cos\theta^{*}$ to the cross section.
We refer to these samples as the ``reweighted \textsc{powheg} MC samples".
The second category contains a number of benchmark beyond-SM (BSM) scenarios generated with the LO generator \textsc{madgraph}~\cite{Alwall:2007st}. These include models containing a $t$-channel $Z'$ boson with a mass of $200~\gevcc$~\cite{PhysRevD.83.114039} or a $s$-channel gluon with an axial coupling (axigluon) with various properties. The axigluon scenarios we simulate include a model with an axigluon near the $\ttbar$ production threshold with pure axial coupling and mass of $425~\gevcc$ (425~GeV Axi)~\cite{PhysRevD.84.054008}, three models with light axigluons with left-handed, pure axial, and right-handed couplings and mass of $200~\gevcc$ (200~GeV AxiL/A/R)~\cite{PhysRevD.87.034039}, and two models with heavy axigluons with a pure axial coupling and masses of $1.8~\text{and}~2.0~\mathrm{TeV/c^{2}}$ (1.8/2.0~TeV Axi)~\cite{cdf_afb_prd2011}.

\section{\label{sec:TopReco}Top-Quark-Pair Reconstruction}

Since the primary goal is to measure the asymmetry as defined in Eq.~(\ref{eq:afbtt}) using $\dy$, we need to reconstruct the kinematic properties of top quark and antiquark on an event-by-event basis. This is achieved by combining the final-state decay products together to form first two $W$-boson candidates and then two top-quark candidates.  This involves pairing each charged lepton with a fraction of the $\met$, corresponding to the momentum of a neutrino, to reconstruct a $W$ boson, and then pairing each resulting $W$ boson with one of the jets to form a top quark. The primary challenges of the reconstruction are to choose the correct lepton-jet pairing, to solve for the neutrino momentum within each pairing, and to determine the best $\ttbar$ kinematic solution when multiple solutions exist. 

We use a likelihood-based algorithm to reconstruct the momenta of the top quarks and antiquarks. 
We sample the kinematically allowed space to obtain the likelihood function of the data.
This information is used to estimate the parton-level results as described in the next section. Additional event-selection criteria, partially based on the reconstruction likelihoods, are used to optimize the sensitivity of the analysis by rejecting poorly-reconstructed top-quark pairs, as well as rejecting non-top-quark-pair events.

In order to determine the four-momenta of both the top quark and antiquark, we need to solve for the four-momenta of all signal decay products. In addition to the individual measurements of charged-lepton and jet momenta, and $\MetVec$, we have additional constraints by using the known masses of the top quark and the $W$ boson in the energy-momentum conservation equations,
\begin{widetext}
    \begin{equation}
        \begin{aligned}
            M_{\ell^{+}\nu}^{2} &=
            (E_{\ell^{+}}+E_{\nu})^2-(\vv{p_{\ell^{+}}}+\vv{p_{\nu}})^2=M_{W}^{2}\\
            M_{\ell^{-}\bar{\nu}}^{2} &=
            (E_{\ell^{-}}+E_{\bar{\nu}})^2-(\vv{p_{\ell^{-}}}+\vv{p_{\bar{\nu}}})^2=M_{W}^{2}\\
            M_{\ell^{+}\nu b}^{2} &=
            (E_{\ell^{+}}+E_{\nu}+E_{b})^2-(\vv{p_{\ell^{+}}}+\vv{p_{\nu}}+\vv{p_{b}})^2=M_{t}^{2}\\
            M_{\ell^{-}\bar{\nu} \bar{b}}^{2} &=
            (E_{\ell^{-}}+E_{\bar{\nu}}+E_{\bar{b}})^2-(\vv{p_{\ell^{-}}}+
            \vv{p_{\bar{\nu}}}+\vv{p_{\bar{b}}})^2=M_{t}^{2}\\
            (\vv{p_{\nu}}+\vv{p_{\bar{\nu}}})_{x} &= \Mex\\
            (\vv{p_{\nu}}+\vv{p_{\bar{\nu}}})_{y} &= \Mey,
        \end{aligned}
        \label{eq:MonEnCons}
    \end{equation}
\end{widetext}
where $x$ and $y$ are the horizontal and vertical coordinates perpendicular to the proton beamline, $z$. The basic ideas and assumptions associated with the top-quark-pair reconstruction used in this analysis are the following:
\begin{enumerate}
    \item Because charged leptons are measured with high precision~\cite{Aaltonen:2014eva}, we neglect resolution effects and assume that their true momenta are the observed momenta.
    \item Because the bottom quarks in this analysis come from the heavy top quarks, the two jets with the largest $\et$ (and $|\eta|<2.5$) are assumed to come from the hadronization of the $b$ and the $\bar{b}$ quarks. The directions of the jets are assumed to correctly indicate the directions of their original quarks. The jet $\et$ values, which are subject to standard corrections~\cite{Bhatti2006375}, are further corrected so that the mean of the difference between the jet $\et$ value and the corresponding $b$-quark $\et$ value is zero~\cite{PhysRevD.73.112006} as estimated from \textsc{powheg} MC samples of $\ttbar$ events. In the reconstruction, the jet $\et$ values are allowed to float around their mean values according the expected resolutions. In addition, we fix the masses of the $b$ quarks to be $4.66~\gevcc$~\cite{Agashe:2014kda}.
    \item Each charged lepton needs to be paired with a $b$ or a $\bar{b}$ quark to form a $t$ or a $\bar{t}$ quark, respectively, together with the neutrinos, which cannot be detected. Since no accurate method is available to separate on an event-by-event basis jets from $b$ quarks and jets from $\bar{b}$ quarks, we consider both lepton-jet pairings in the reconstruction, and use techniques described later in this section to statistically reduce the contamination of the measurement from wrong pairings.
    \item While the two neutrinos in the final state are not detected, resulting in six unknown variables assuming massless neutrinos, the sum of the transverse momenta of the two neutrinos produces the $\MetVec$ in the event~\cite{MET}. Since the two measured components of $\MetVec$ ($\Mex$ and $\Mey$) have large uncertainties, in the reconstruction the vector sum of the transverse momenta of the neutrinos is allowed to float around the measured central value according to its resolution.
    \item In all calculations we assume that the $W$ bosons and the top quarks are on mass shell, thus including four constraints in the $\ttbar$ system: the two $W$-boson masses ($m_{W} = 80.4~\gevcc$) and the two top-quark masses ($m_{t} = 172.5~\gevcc$)~\cite{Agashe:2014kda}. The systematic uncertainty due to the uncertainties on the assumed masses is negligible.

\end{enumerate}

With these assumptions, within each of the two lepton-jet pairings, there are ten unknown variables in the $\ttbar$ dilepton final state, six from the momenta of the two neutrinos, two from the floating jet-$\et$ values and two from the floating components of $\MetVec$. On the other hand, we have six constraints from Eq.~(\ref{eq:MonEnCons}). Thus, for each event there are two underconstrained systems with multiple solutions in a four-dimensional parameter space. We scan these two four-dimensional parameter spaces and assign a likelihood to each point of the phase space based on the measured quantities and their uncertainties. In the next paragraph, additional information about the expected $p_{z}$, $\pt$, and invariant mass of the $\ttbar$ system ($\pztt$, $\pttt$, and $\Mtt$, respectively) is also incorporated into the likelihood to improve the reconstruction. By incorporating this information, we are assuming that the $\pztt$, $\pttt$, and $\Mtt$ spectra follow the predictions of SM at NLO. The results of this analysis need to be interpreted under this assumption. Any bias caused by this assumption is discussed in Sec.~\ref{sec:validation}.

With these sets of assumptions, the kinematic properties of a $\ttbar$ event are characterized as functions of the momenta of the neutrinos ($\vv{p_{\nu}}$ and $\vv{p_{ \bar{\nu}}}$) and the transverse energy of the $b$ and $\bar{b}$ quarks ($E_{T,b}$ and $E_{T,\bar{b}}$). The quantities $\vv{p_{\nu}}$, $\vv{p_{ \bar{\nu}}}$, $E_{T,b}$, and $E_{T,\bar{b}}$ are not independent of each other, but are subject to the constraints of Eq.~(\ref{eq:MonEnCons}). In the kinematically allowed region, we define the following likelihood to quantify the goodness of a solution:
\begin{widetext}
    \begin{equation}
        \begin{aligned}
            \mathcal{L}&(\vv{p_{\nu}}, \vv{p_{ \bar{\nu}}}, E_{T,b}, E_{T,\bar{b}}) = P(\pztt)\times P(\pttt)\times            P(\Mtt)\\
                       &\times \frac{1}{\sigma_{\mathrm{jet1}}}\exp\left[ -\frac{1}{2}\left(\frac{E_{T, \mathrm{jet1}}-E_{T, b}}{\sigma_{\mathrm{jet1}}}\right)^{2}\right] \times
            \frac{1}{\sigma_{\mathrm{jet2}}}\exp\left[ -\frac{1}{2}\left(\frac{E_{T, \mathrm{jet2}}-E_{T, \bar{b}}}{\sigma_{\mathrm{jet2}}}\right)^{2}\right]\\
            &\times \frac{1}{\sigma({\Mex})}\exp\left[ -\frac{1}{2}\left(\frac{\Mex-(\vv{p_{\nu}}+\vv{p_{\bar{\nu}}})_{x}}{\sigma(\Mex)}\right)^{2}\right]
            \times \frac{1}{\sigma(\Mey)}\exp\left[ -\frac{1}{2}\left(\frac{\Mey-(\vv{p_{\nu}}+\vv{p_{\bar{\nu}}})_{y}}{\sigma(\Mey)}\right)^{2}\right],
        \end{aligned}
        \label{eq:recoLikelihood}
    \end{equation}
\end{widetext}
where $P(\pztt)$, $P(\pttt)$, and $P(\Mtt)$ are the probability-density functions of each parameter obtained from the simulated $\ttbar$ signal events that pass the selection requirements, the two $E_{T, \mathrm{jet}}$ values are the measured transverse energies of the two jets, the two $\sigma_{\mathrm{jet}}$ values are the expected resolutions of the jet transverse energies estimated with the same signal sample, $\Mexy$ are the $x$ and $y$ components of the measured $\MetVec$, and $\sigma(\Mexy)$ are the expected resolutions of $\Mexy$, estimated with the same sample. The parameters $\left( E_{T, \mathrm{jet1,2}}-E_{T, b,\bar{b}}\right)/\left(\sigma_{\mathrm{jet1,2}}\right)$ quantify the deviation between the hypothetical $b$-quark $\et$ values and the measured jet $\et$ values, and are referred to as ``jet deviations" ($\delta_{j,1}$ and $\delta_{j,2}$), where the labeling of 1 and 2 is random.

\begin{figure*}[hbtp]
    \includegraphics[width=\linewidth]{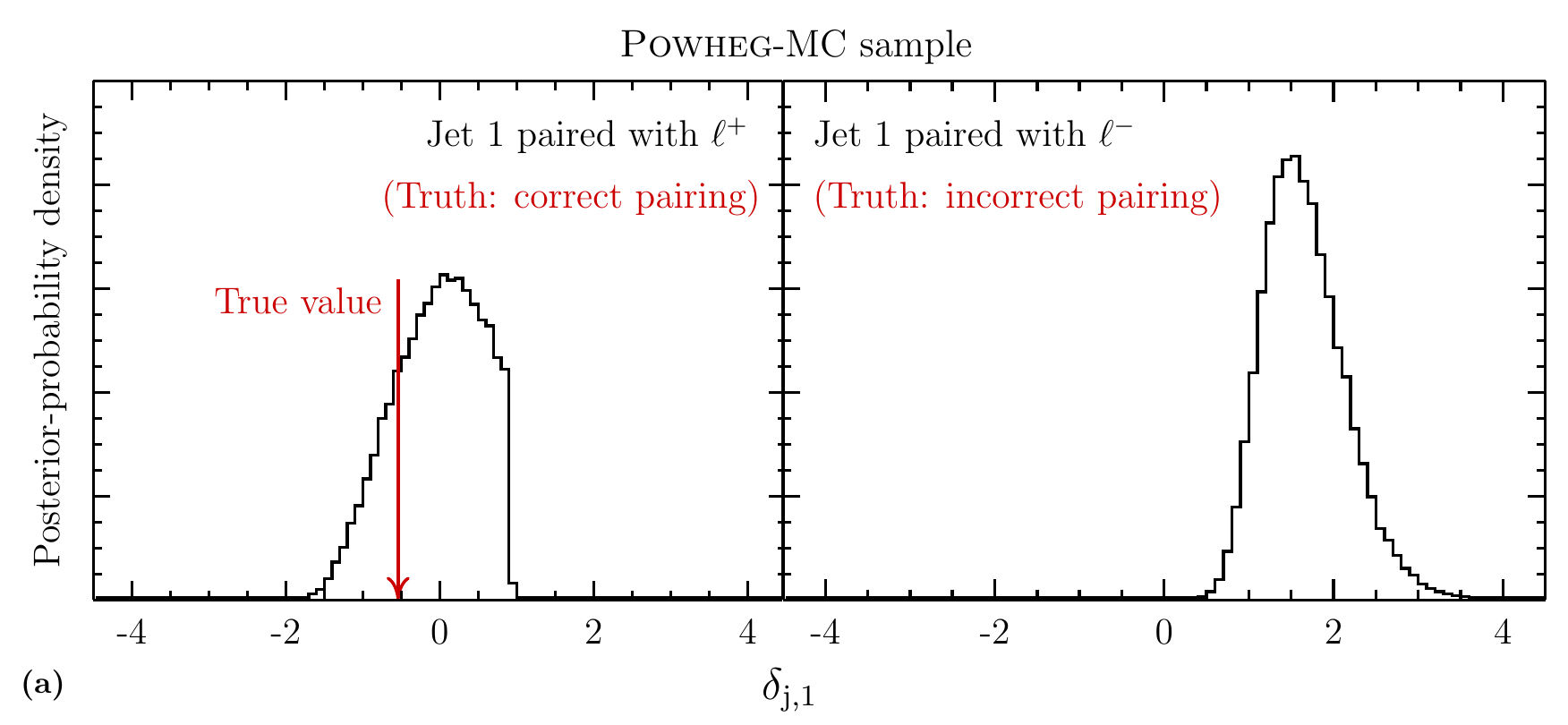}\\
    \includegraphics[width=\linewidth]{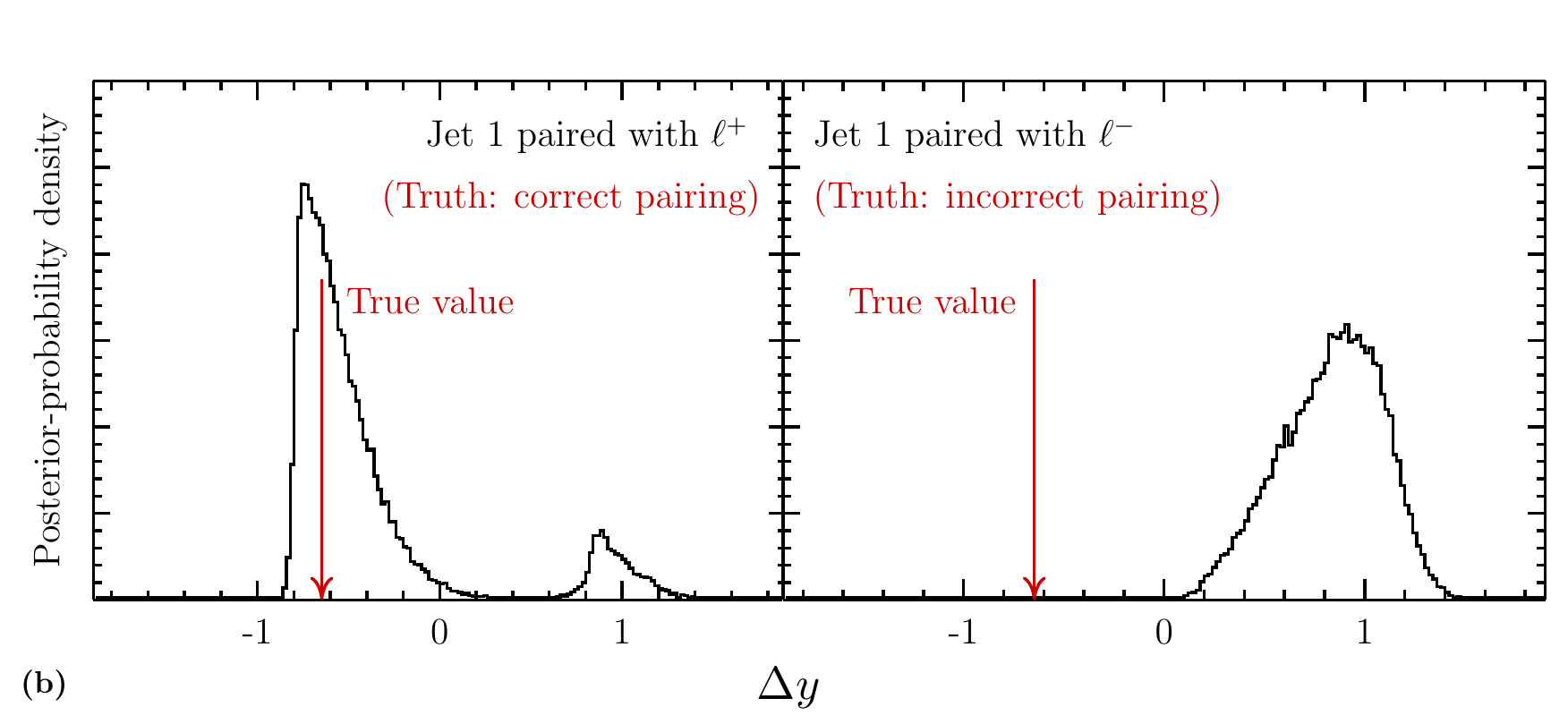}
    \caption{Posterior-probability density of $\jd$ for one of the two jets (a) and $\dy$ (b) for one example event from the \textsc{powheg}-MC sample of $\ttbar$ events. Based on the generator-level information, the left panels refer to the correct lepton-jet pairing and the right panels refer to the incorrect pairing. The red vertical arrows show the true values of the parameters.}
    \label{fig:RecoExample}
\end{figure*}

We employ a Markov-chain Monte Carlo (MCMC) method~\cite{Metropolis} to efficiently sample from the kinematic parameter space with each of the two lepton-jet pairings. The probability distribution of $\dy$ is obtained by marginalizing over the distributions of all other parameters~\cite{Caldwell20092197}. 
An example probability distribution for one of the two $\jd$ parameters and the $\dy$ parameter for one event from the \textsc{powheg} signal sample is shown in Fig.~\ref{fig:RecoExample}. Based on the information from the event generator, the left panels refer to the correct lepton-jet pairing and the right panels refer to the incorrect pairing. The vertical arrows show the true values of the parameter.
To make the best use of the available information, we use the probability-density distributions obtained from the MCMC method in the extraction of the parton-level asymmetry, and weight the two lepton-jet pairings based on the maximum likelihood achieved in each of the two pairings ($L_{\text{max}, 1, 2}$). The weight of each lepton-jet pairing is determined by
\begin{equation}
    w_{1, 2} = \frac{L_{\text{max}, 1, 2}}{L_{\text{max}, 1}+L_{\text{max}, 2}},
    \label{eq:EvenOddWeight}
\end{equation}
so that the total weight $w_1+w_2$ of each event is unity. The information used in the parton-level $\afbtt$ extraction comes from the sum of the $\dy$ distributions of the two lepton-jet pairings weighted by Eq.~(\ref{eq:EvenOddWeight}).
With this set of choices we find that the resolution of the top-quark-pair reconstruction algorithm is approximately 0.5 in $\dy$. 

\section{\label{sec:Unfolding}Extracting parton-level asymmetry}

We introduce a Bayesian procedure to extract a parton-level measurement of $\afbtt$ from the $\dy$ distribution observed in data. The two differ because of the limited acceptance and efficiency of the detector, imperfect resolution of the $\dy$ reconstruction, and the background contributions. The procedure takes into account correlations between measured $\dy$ values and allows a determination of both the inclusive and differential asymmetries. 
A $4\times4$ matrix models the reconstruction resolution, by mapping the parton-level $\dy$ into the reconstructed $\dy$. The edges $-\infty$, $-0.5$, $0$, $0.5$, and $\infty$ of bin 1, 2, 3, and 4, respectively, are chosen to preserve approximately equal numbers of expected events in each bin after reconstruction. The forward region consists of bin 3, and 4, while the backward region consists of bin 1, and 2.
The parton-level inclusive $\afbtt$ is expressed as
\begin{widetext}
    \begin{equation}
        \afbtt=\frac{\nparton[3]+\nparton[4]-\nparton[1]-\nparton[2]}{\nparton[3]+\nparton[4]+\nparton[1]+\nparton[2]},
        \label{eq:trueAFB}
    \end{equation}
\end{widetext}
where $\nparton[p]$ represents the hypothesized parton-level event rate in the $p$th bin. The expected number of events in the $r$th bin after the top-quark-pair reconstruction for a particular set of $\nparton[p]$ is expressed as
\begin{equation}
    \nexp[r]=\sum\limits_{p=1}^{4} \nparton[p]\cdot\eff[p](\afbtt)\cdot\detS[p][r]+\nbkg[r],
    \label{eq:Unfolding}
\end{equation}
where
\begin{itemize}
    \item $\eff[p](\afbtt)$ represents the efficiency in the $p$th bin at parton level, which accounts for the acceptance imposed by the detector coverage and the efficiency associated with the event selection, which is a function of the parton-level value of $\afbtt$,
    \item $\detS[p][r]$ represents the smearing matrix, which is symmetric and accounts for the detector resolution and the smearing caused by the top-quark-pair reconstruction procedure, is observed not to change as a function of the input $\afbtt$, and
    \item $\nbkg[r]$ is the expected background contribution of the $r$th bin.
\end{itemize}
The $\eff[p](\afbtt)$ term is estimated with the reweighted \textsc{powheg} MC samples described in Sec.~\ref{sec:TOPDILEventSelection}, and the $\detS[p][r]$ term is estimated with the nominal \textsc{powheg} MC sample and normalized so that $\sum_p \detS[p][r] = 1$.
The observed bin count from data $\nobs[r]$ is compared with the expectation $\nexp[r]$ with a $\chi^{2}$ fit, with correlations among bins estimated with the \textsc{powheg} $\ttbar$ MC sample. 

To allow for the use of well-motivated prior probability distributions, we reparametrize the hypothesized $\nparton[p]$ as follows:
\begin{enumerate}
    \item $N_{\text{tot}}=\sum\limits_{p=1}^{4} \nparton[p]$ is the total number of signal events, with a uniform prior probability distribution in $(0,\infty)$
    \item $A_{\text{in}} =  \frac{\nparton[3]-\nparton[2]}{\nparton[3]+\nparton[2]} = \afbtt(|\dy|<0.5)$ is the asymmetry of bins 2 and 3, with a uniform prior in (-1, 1)
    \item $A_{\text{out}} =  \frac{\nparton[4]-\nparton[1]}{\nparton[4]+\nparton[1]} = \afbtt(|\dy|>0.5)$ is the asymmetry of bins 1 and 4, with a uniform prior in (-1, 1)
    \item $R_{\text{in}} =  \frac{\nparton[2]+\nparton[3]}{N_{\text{tot}}}$ is the fraction of events in the inner two bins, with a uniform prior in (0, 1).
\end{enumerate}
The prior-probability distributions of $A_{\text{out}}$ and $A_{\text{in}}$ are assigned to be uniform to assume no knowledge on these parameters. The final result is not sensitive to the prior probability distribution of $R_{\text{in}}$. 
With this new parametrization, the inclusive $\afbtt$ in Eq.~(\ref{eq:trueAFB}) is written as
\begin{equation}
    \afbtt = R_{\text{in}}A_{\text{in}}+(1-R_{\text{in}})A_{\text{out}}.
    \label{eq:AFBttRepar}
\end{equation}

The posterior-probability distribution of each parameter of interest ($\afbtt$, $\afbtt(|\dy|<0.5)$ and $\afbtt(|\dy|>0.5)$) is obtained by marginalizing over the distributions of all other parameters. The measured values of the parameters and their statistical uncertainties are extracted by fitting a Gaussian function to the core of the resulting posterior distribution of the parameter of interest.

The procedure is validated and the uncertainties are estimated using two ensembles of 5,000 pseudoexperiments each. One set of pseudoexperiments is generated by randomly sampling events from the nominal \textsc{powheg} MC sample with the number of events following the signal expectation for data. The second set is generated by randomly sampling events from both the signal and the background estimation samples in the same way. The parton-level $\afbtt$ is estimated in each pseudoexperiment using the procedure described above. The pseudoexperiments are used to test for potential biases as well as to determine the expected statistical uncertainty with signal only, and the total statistical uncertainty when the backgrounds are included. As is shown in Sec.~\ref{sec:validation}, no bias is observed. 
The expected total statistical uncertainty for the inclusive measurement in data is estimated as the standard deviation of the results from the second set of pseudoexperiments.
Before the optimization we describe in Sec.~\ref{sec:optimization}, this expected total statistical uncertainty is around 0.12, and is expected to be the dominant uncertainty. As in Ref.~\cite{Aaltonen:2014eva}, we take the systematic uncertainty due to the uncertainty on the background normalization and shape to be equal to the difference, in quadrature, between the total statistical uncertainty and the signal-only statistical uncertainty, as it captures the uncertainty caused due to the existence of the background. The background systematic uncertainty is estimated to be 0.06 before the optimization.
Additional uncertainties are described in Sec.~\ref{sec:Systematics}.

\section{\label{sec:optimization}Optimization}
We implement an optimization procedure to improve the top-quark-pair reconstruction and asymmetry determination.
The goal of the optimization is to minimize the quadrature sum of the expected total statistical uncertainty and the background systematic uncertainty, as other uncertainties are expected to be small. 
Besides the statistical uncertainty due to the limited data sample size, the uncertainty of the parton-level $\afbtt$ measurement receives a contribution from the resolution of the reconstructed $\dy$. This contribution is dominated by events in which the reconstructed value of $\dy$ differs significantly from the true parton-level value. Reducing this fraction of poorly reconstructed events effectively reduces the uncertainty of the measurement. The usual reconstruction method of only using the solution that maximizes the likelihood~\cite{Aaltonen:2012it} suffers from two primary problems: 1) the algorithm occasionally selects the wrong lepton-jet pairing and 2) the algorithm occasionally gives the highest likelihood values to a set of solutions to Eq.~(\ref{eq:MonEnCons}) that is different from the one corresponding to the real event within the right lepton-jet pairing. To ameliorate these problems we calculate the probability distributions associated to both options of lepton-jet pairings and use them to calculate weights instead of choosing the maximum-likelihood solution. This improves the resolution for the $\afbtt$ measurement, as it reduces the expected statistical uncertainty of the inclusive $\afbtt$ by approximately $15\%$ (relative).

We further optimize by incorporating additional selection requirements to reject badly-reconstructed lepton-jet pairings and by giving larger weights to pairings that are more likely to be the correct ones. 
For wrong lepton-jet pairings or background events, the top-quark-pair reconstruction algorithm occasionally yields a heavily biased estimate of $\et$ to try to make a valid $\ttbar$ pair, resulting in a large $|\jd|$. For simplicity we examine only the maximum, $\jdpeak$, of the posterior-probability distribution of $\jd$ for each jet. We reject any lepton-jet pairing with $\sqrt{\delta_{\text{j,peak1}}^{2}+
\delta_{\text{j,peak2}}^{2}}>\Theta(\jd)$, where $\Theta(\jd)$ is the threshold to be optimized, and reject the
event if both lepton-jet pairings are rejected.

The jet charge $Q_{\mathrm{jet 1,2}}$, calculated with the \textsc{JetQ} algorithm~\cite{PhysRevD.88.032003}, is correlated with the charge of the quark that originated the jet and provides additional separation between the $b$ quark and the $\bar{b}$ quark, thus helping to identify the correct lepton-jet pairing. This technique was recently used in Ref.~\cite{Aaltonen:2015mba}. While the jet charge suffers from dilution due to bottom-hadron oscillations and cascade decays, and biases due to the detector material and track reconstruction~\cite{Abazov:2014lha}, it still provides a worthwhile improvement in the resolution of $\afbtt$. For each event we examine the sign of $\Delta Q = Q_{\mathrm{jet1}}-Q_{\mathrm{jet2}}$, where the labeling of 1 and 2 is random, after assigning $Q_{\mathrm{jet}}=0$ for jets without valid reconstructed charges for simplicity; positive values of $\Delta Q$ suggest that jet1 is from the $\bar{b}$ quark and jet2 is from the $b$ quark, and vice versa. The case $\Delta Q=0$ indicates that the jet charge is unable to provide distinguishing power. To use this information, we introduce a global jet-charge probability weight $w_{Q}$ that quantifies the probability that the jet charge gives the correct lepton-jet pairing. We then amend the $L_{\text{max}}$ of the two pairings used in Eq.~(\ref{eq:EvenOddWeight}) to $L_{\text{max}}\times w_{Q}$ if $\Delta Q$ suggests this pairing and $L_{\text{max}}\times (1-w_{Q})$ if $\Delta Q$ suggests otherwise, and proceed with Eq.~(\ref{eq:EvenOddWeight}) in determining the weights of the two pairings. We optimize for the value of $w_{Q}$. 

A third improvement consists in rejecting the lepton-jet pairings with high $m_{lb}^{2}$, which are unlikely to originate from a top-quark decay, where $m_{lb}$ is the invariant mass of the lepton+$b$-quark system~\cite{mlbJA}. We reject any lepton-jet pairings with $m_{lb}^{2} > \Theta(m_{lb}^{2})$, and reject the event if both lepton-jet pairings are rejected. We optimize for the value of $\Theta(m_{lb}^{2})$. 

Finally, events with a lepton appearing too close to a jet either cannot be well reconstructed or are likely to result from associated production of a $W$ boson and a jet where a $b$ jet is reconstructed both as a lepton and a jet~\cite{Aaltonen:2015hta}, which happens when a muon is present in the b-quark hadronization process. This effect is quantified using the minimum radius $\Delta R=\sqrt{(\Delta\eta)^{2}+(\Delta\phi)^{2}}$ between any lepton and any jet ($\Delta R_{\text{min}}(\ell,j)$). We optimize for a requirement of $\Delta R_{\text{min}}(\ell,j)>\Theta(\Delta R_{\text{min}})$ as it helps reject $W$+jets background events without significantly reducing the number of well-reconstructed $\ttbar$ events.

\begin{figure*}[htbp]
    \includegraphics[width=\linewidth]{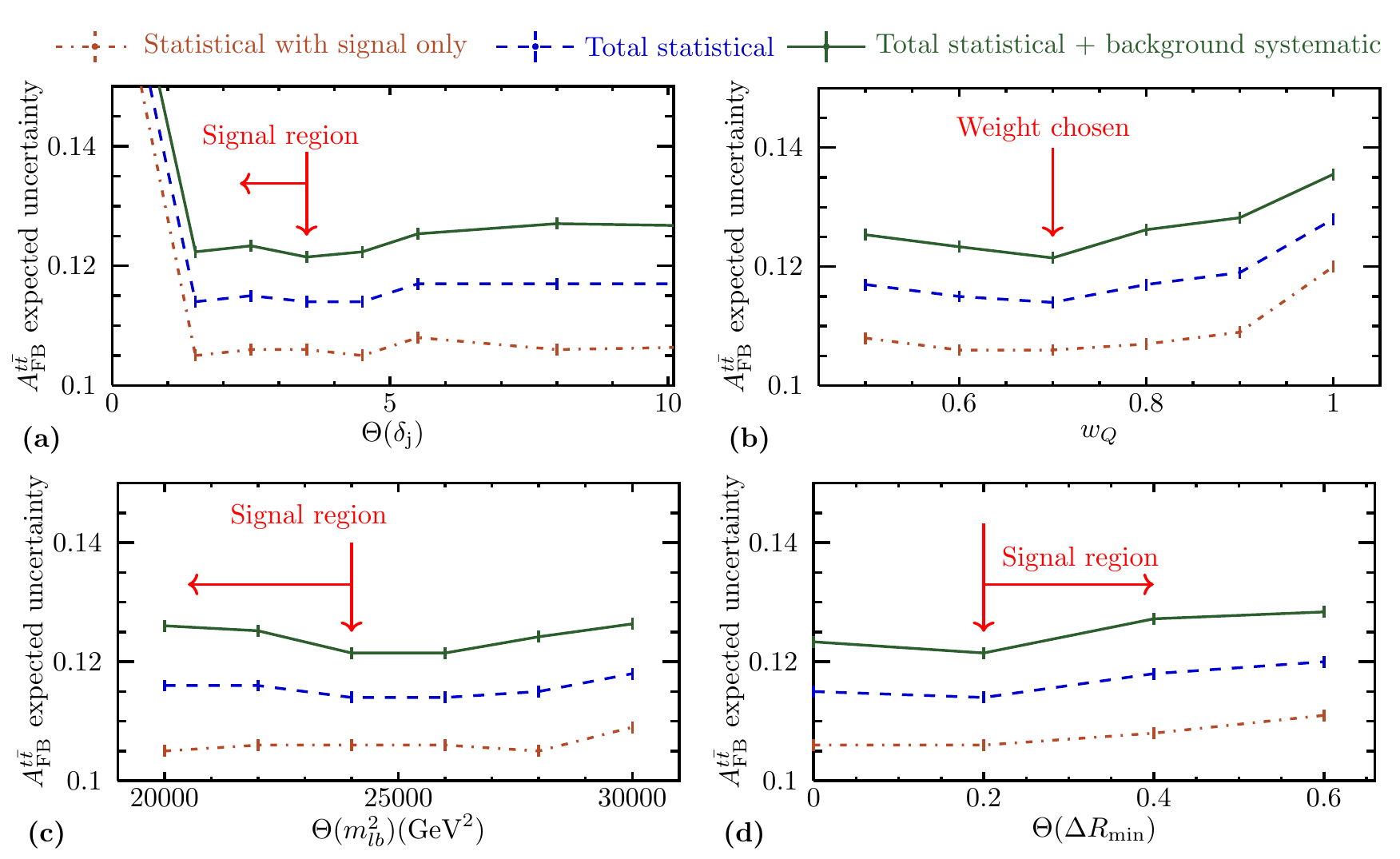}
    \caption{Expected uncertainties as functions of the four optimization parameters. In each plot is shown the statistical uncertainty for signal only (dash-dotted line), statistical uncertainty for signal and backgrounds (total statistical uncertainty, dashed line) and the quadrature sum of the total statistical uncertainty and the background systematic uncertainty (solid line). The optimal values are based on the minimum point of the green solid line, as marked with the vertical arrows on the plots, and summarized in Table~\ref{tab:OptiSummary}. For each plot, all other optimization parameters are held at their optimal values.}
    \label{fig:OptiResult}
\end{figure*}

The minimizations of the expected uncertainty for all criteria and weight values are done simultaneously. Table~\ref{tab:OptiSummary} shows the optimal values. Figure~\ref{fig:OptiResult} shows the expected uncertainties as functions of the criteria and weights with other values fixed at the optimal points. We proceed with the analysis with this optimized configuration. The resolution in $\dy$ after the optimization remains approximately 0.5. The signal efficiency of the top-quark-reconstruction requirements is 95\% with a background rejection of 40\% relative to the baseline event selection requirements. The minimum expected uncertainties achieved are 0.106 for the signal-only statistical uncertainty, 0.114 for the statistical uncertainty of signal and backgrounds (total statistical uncertainty, improved by 7\%), and 0.121 for the quadrature sum of the total statistical and the background systematic uncertainties (improved by 11\%). For the differential measurement, we find expected total statistical uncertainties of 0.34 for $\afbtt(|\dy|<0.5)$ and 0.16 for $\afbtt(|\dy|>0.5)$.

\begin{table}[hbtp]
    \caption{Summary of the criteria and weight values used to optimize the expected uncertainties in the measurement of the inclusive $\afbtt$.}
    \label{tab:OptiSummary}
    \TableCutWeightOptimal
\end{table}

\section{\label{sec:validation}Validation}

\begin{figure*}[htbp]
    \includegraphics[width=\linewidth]{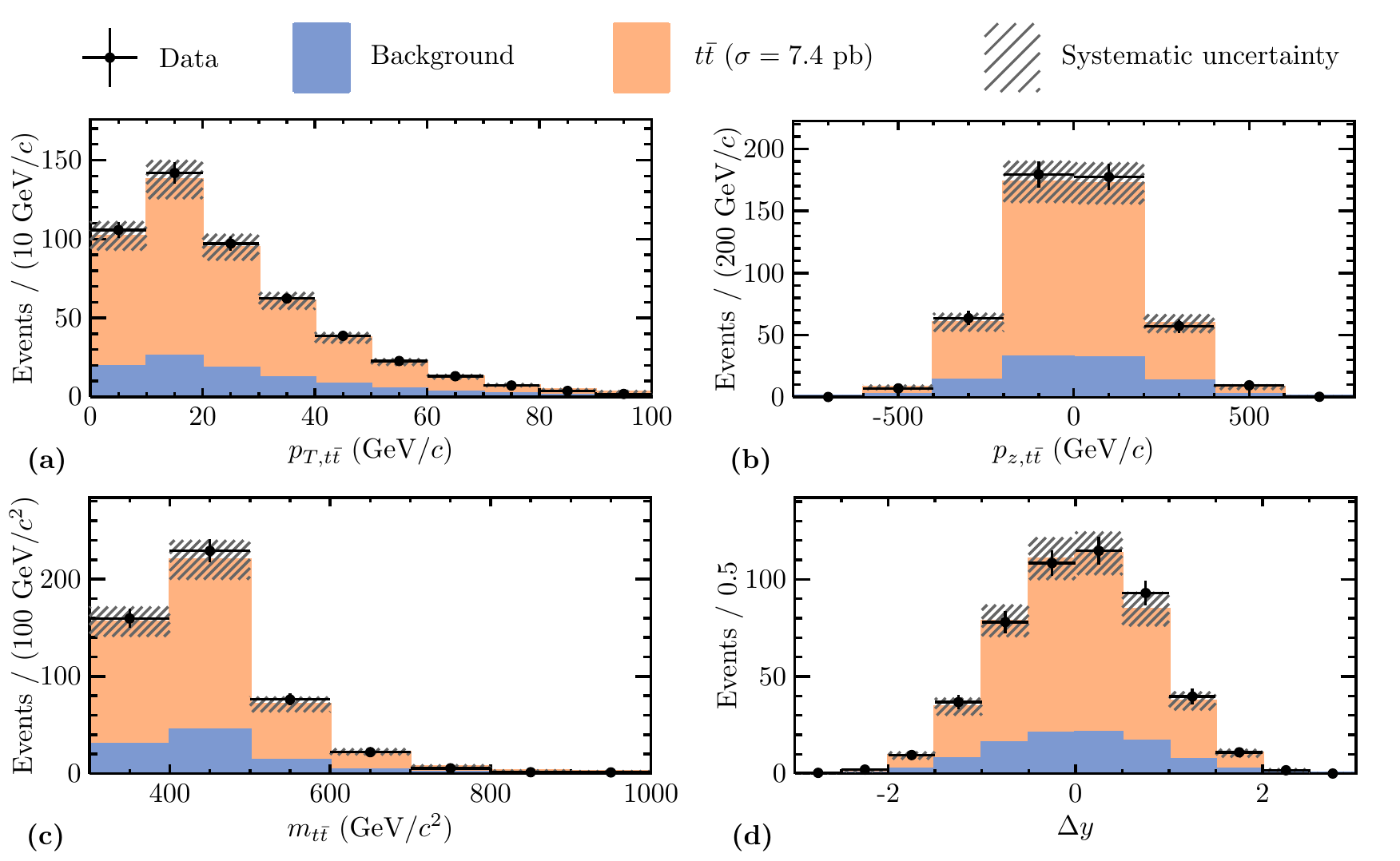}
    \caption{Distributions of $\pttt$ (a), $\pztt$ (b), $\Mtt$ (c), and $\dy$ (d) from data compared with the SM expectations.}
    \label{fig:KinValidTopReco}\label{fig:dyData}
\end{figure*}

The expected numbers of events from all SM sources, along with the observed number of events passing all the baseline event selections and the top-quark-pair reconstruction quality selections, are summarized in Table~\ref{tab:EventYieldOpt}. 
The distributions of $\pttt$, $\pztt$, and $\Mtt$ from data are shown in Fig.~\ref{fig:KinValidTopReco}(a)-(c) and compared to the signal and background models. The agreement between data and the predictions is good. The distribution of reconstructed $\dy$ is shown in Fig.~\ref{fig:dyData}(d). The $\afbtt$ result is extracted from this distribution.

\begin{table}[hbtp]
    \caption{Expected and observed number of events passing all the baseline event selections and the top-quark-pair reconstruction quality selections. The quoted uncertainties are the quadratic sums of the statistical and systematic uncertainties in each row.}
    \label{tab:EventYieldOpt}
    \TableEventYield
\end{table}

\begin{figure}[hbtp]
    \includegraphics[width=\columnwidthfigure]{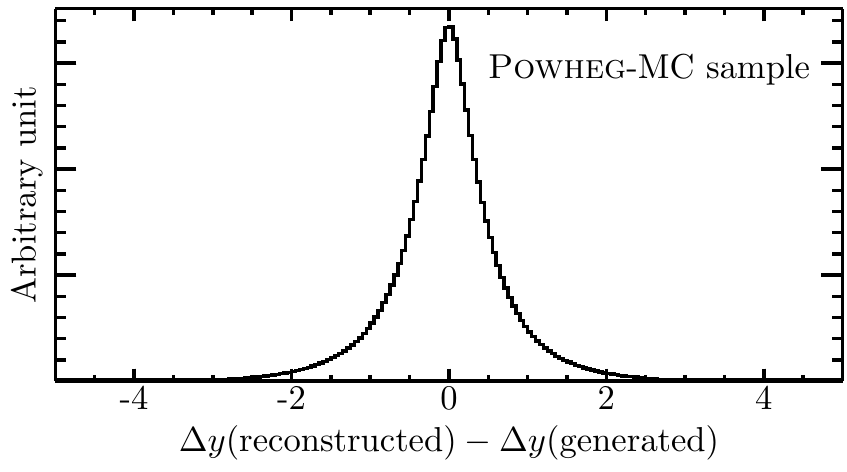}
    \caption{Distribution of the difference between reconstructed and generated values for $\dy$ from events in the nominal \textsc{powheg} $\ttbar$ MC after all the event-selection criteria. Each event contributes a probability distribution with a unity weight. \label{fig:DeltaDeltaOpt}}
\end{figure}

\begin{figure}[hbtp]
    \includegraphics[width=\columnwidthfigure]{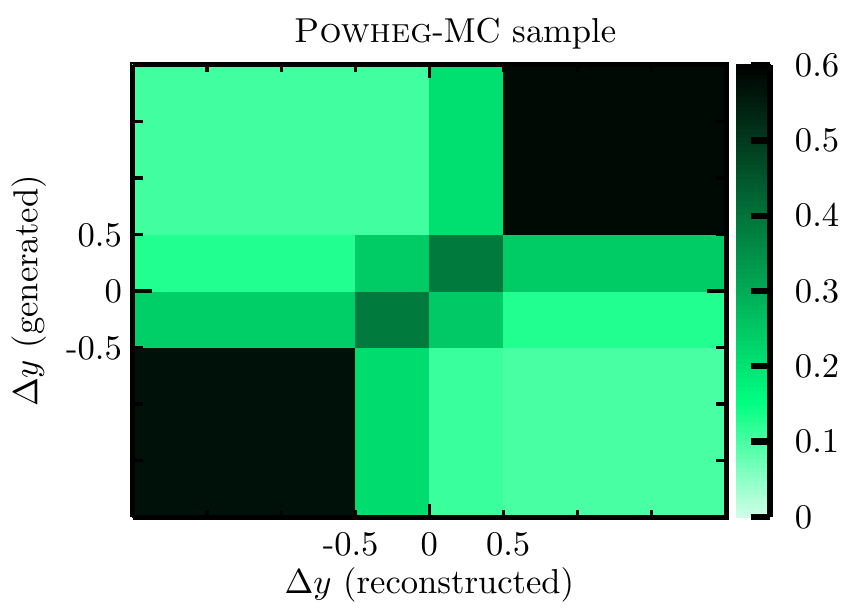}
    \caption{Detector smearing matrix estimated with the nominal \textsc{powheg} $\ttbar$ MC sample.\label{fig:DetS_opt}}
\end{figure}

\begin{figure}[htbp]
    \includegraphics[width=\columnwidthfigure]{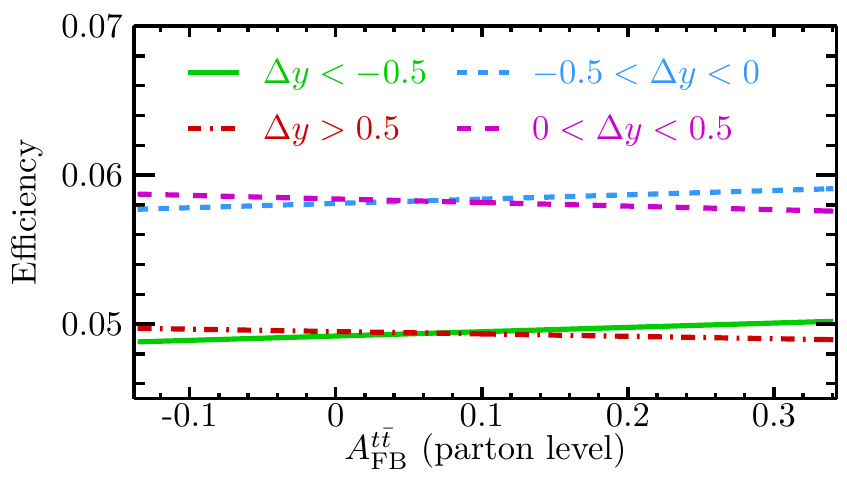}
    \caption{Efficiencies in the four bins, approximated to linear functions of the $\afbtt$, estimated with the reweighted \textsc{powheg} MC samples.}
    \label{fig:EffSum}
\end{figure}

Figure~\ref{fig:DeltaDeltaOpt} shows the reconstruction resolution, defined as the difference between reconstructed and generated values of $\dy$, estimated for events from the \textsc{powheg} MC samples. The distribution shown in this figure is obtained by summing the posterior-probability distribution of the reconstruction resolution over all events in the sample, where each event is weighted equally. In 61\% of the cases the $\dy$ is reconstructed within 0.5 of its true value. The detector smearing matrix $\detS[p][r]$ is shown in Fig.~\ref{fig:DetS_opt}. The efficiencies $\eff[p]$ in the four bins are approximated to linear functions of $\afbtt$ and are shown in Fig.~\ref{fig:EffSum}.

\begin{figure}[htbp]
    \includegraphics[width=\columnwidthfigure]{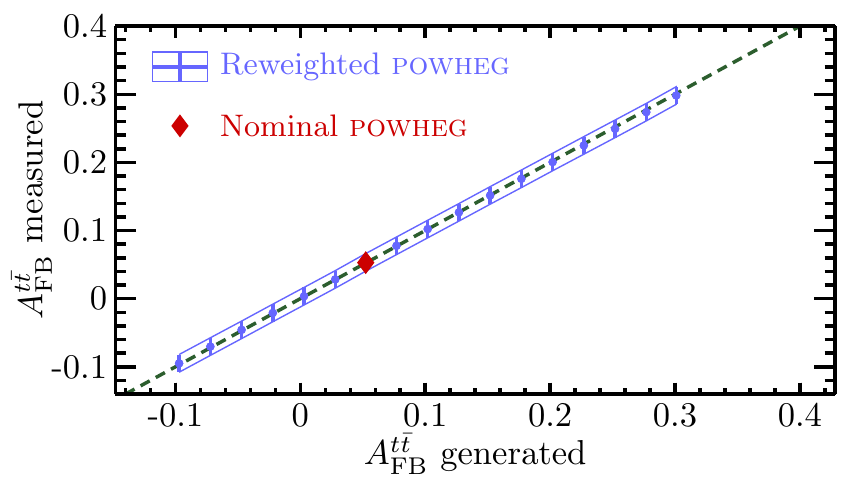}
    \caption{Comparison of the $\afbtt$ values observed in the reweighted \textsc{powheg} MC samples and the $\afbtt$ values generated. The dashed line shows where the measured and generated values are equal. No bias is observed.}
    \label{fig:AFBMAFBGPowheg}
\end{figure}

\begin{figure}
    \includegraphics[width=\columnwidthfigure]{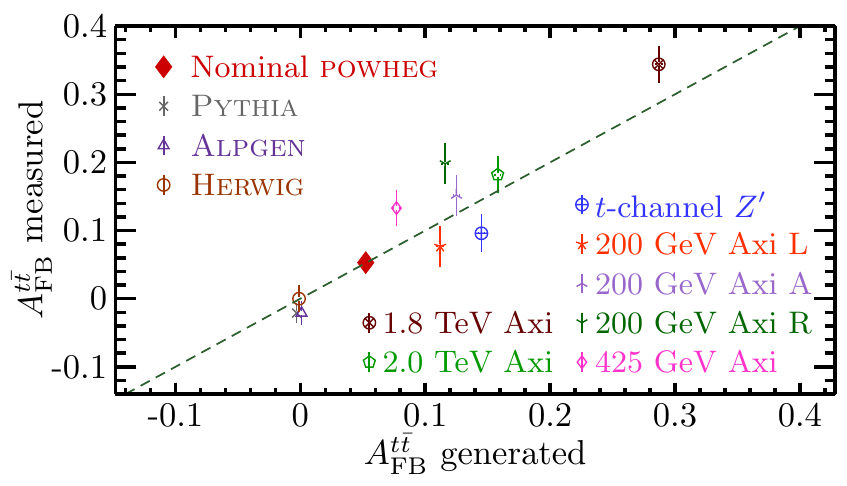}
    \caption{Same as Fig.~\ref{fig:AFBMAFBGPowheg}, but with a number of predicted values for $\afbtt$ from LO SM calculations and a few benchmark BSM scenarios. The description of the BSM scenarios is in the main text.}
    \label{fig:AFBMAFBGAll}
\end{figure}

We test the parton-level $\afbtt$ estimation procedure with the reweighted \textsc{powheg} MC samples. The results are shown in Fig.~\ref{fig:AFBMAFBGPowheg}. The error bars correspond to the statistical uncertainties based on a sample of 70~000 simulated events that meet the selection criteria. No bias is observed.
In addition, we test the estimation procedure with the LO SM calculations from \textsc{pythia}~\cite{Sjostrand:2006za}, \textsc{alpgen}~\cite{Mangano:2002ea}, and \textsc{herwig}~\cite{Bahr:2008pv} as well as a series of benchmark non-SM scenarios described in Sec.~\ref{sec:TOPDILEventSelection}. The results are shown in Fig.~\ref{fig:AFBMAFBGAll}. We do not expect the estimation of $\afbtt$ to be unbiased in all non-SM scenarios, since the assumptions on the $\pztt$, $\pttt$, and $\Mtt$ distributions we made in top-quark-pair reconstruction no longer hold, both due to the effect of non-SM dynamics and because these samples are only calculated at LO. Particularly, the $\pttt$ spectrum calculated at LO shows deviation from data due to lack of higher-order amplitudes with non-zero $\pttt$, while the NLO calculation provides reasonable agreement~\cite{Aaltonen:2012it}. The largest deviation is 0.08.

\begin{figure}[hbtp]
    \includegraphics[width=\columnwidthfigure]{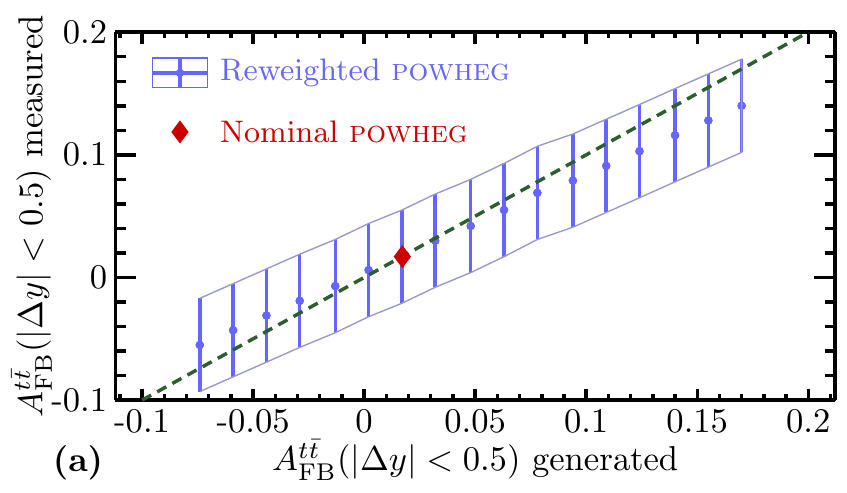}\\
    \includegraphics[width=\columnwidthfigure]{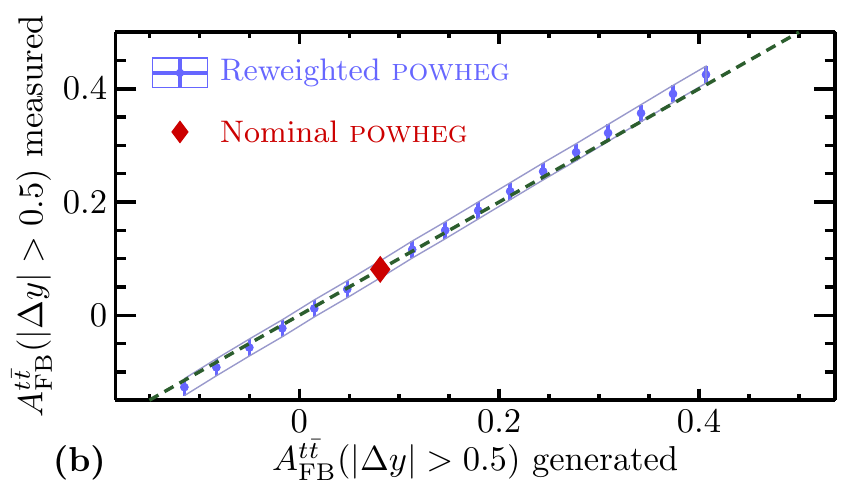}
    \caption{Same as Fig.~\ref{fig:AFBMAFBGPowheg}, but for the $\afbtt(|\dy|<0.5)$ (a) and $\afbtt(|\dy|>0.5)$ (b) measurements. The uncertainties correspond to the size of the \textsc{powheg} MC sample which is over a factor of 100 larger than the data, and the measured values are always within $1\sigma$ of the generated values. (Note the different vertical scales in the two subfigures.)
    }
    \label{fig:AFBMAFBGdY}
\end{figure}

Figure~\ref{fig:AFBMAFBGdY} shows a comparison of $\afbtt (|\dy|<0.5)$ and $\afbtt (|\dy|>0.5)$ between the measured values from the reweighted \textsc{powheg} MC samples and their input values. The error bars correspond to the statistical uncertainties with the entire \textsc{powheg} MC sample which is over a factor of 100 larger than the data. 
The small potential bias shown in Fig.~\ref{fig:AFBMAFBGdY} is negligible compared to the expected statistical uncertainties in the data.
We do not correct for this potential bias and take the difference between the generated and measured asymmetry at the measured central values from data as a systematic uncertainty.

\section{\label{sec:Systematics}Systematic Uncertainties}

In addition to the systematic uncertainty due to the background, several other sources are considered. We estimate the systematic uncertainty due to the potential biases in the NLO SM assumption made in the top-quark-pair reconstruction and in the parton-level asymmetry extraction as the difference between the generated $\afbtt$ and the measured $\afbtt$ with the \textsc{pythia} MC sample. The systematic uncertainty due to the modeling of parton showering and color coherence~\cite{Aaltonen:2012it}, the modeling of color reconnection~\cite{PhysRevD.81.031102}, the amount of initial- and final-state radiation, the size of the jet-energy scale corrections~\cite{Bhatti2006375}, and the underlying parton-distribution functions~\cite{Abulencia:2005aj} are evaluated by repeating the measurement after introducing appropriate variations into the assumptions used in modeling the behavior of the signals, following Ref.~\cite{Aaltonen:2012it}. Table~\ref{tab:systematics} summarizes the statistical and systematic uncertainties of the inclusive $\afbtt$ measurement, and Table~\ref{tab:systematics} summarizes the uncertainties for the $\afbtt$ vs. $|\dy|$ measurements.

\begin{table*}[hbtp]
    \caption{Uncertainties for the inclusive $\afbtt$, $\afbtt(|\dy|<0.5)$ and $\afbtt(|\dy|>0.5)$ measurements.}
    \label{tab:systematics}
    \TableUncertaintyDifferential
\end{table*}

\section{\label{sec:results}Dilepton Results}
\begin{figure}[hbtp]
    \includegraphics[width=\columnwidthfigure]{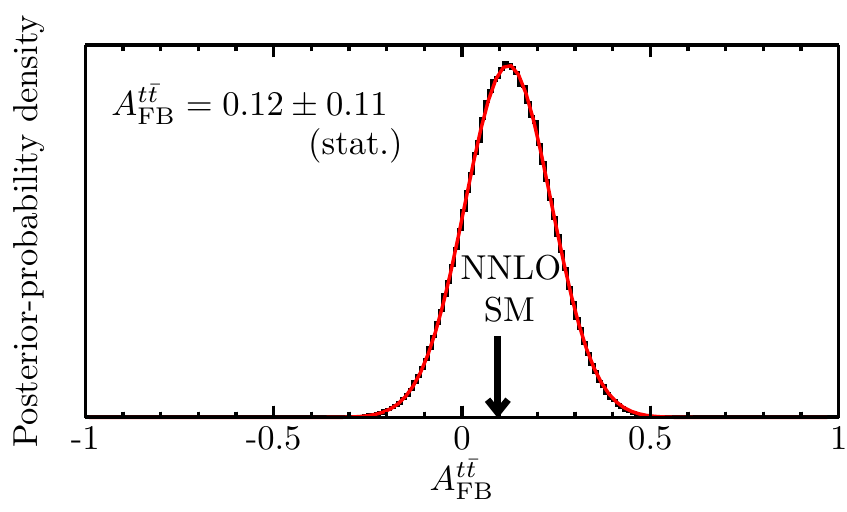}
    \caption{Posterior-probability density for the measurement of the inclusive $\afbtt$. A Gaussian function is fitted to the core of the distribution to extract the result. The NNLO SM prediction is $0.095\pm0.007$.}
    \label{fig:AFBResult}
\end{figure}

\begin{figure}
    \includegraphics[width=\columnwidthfigure]{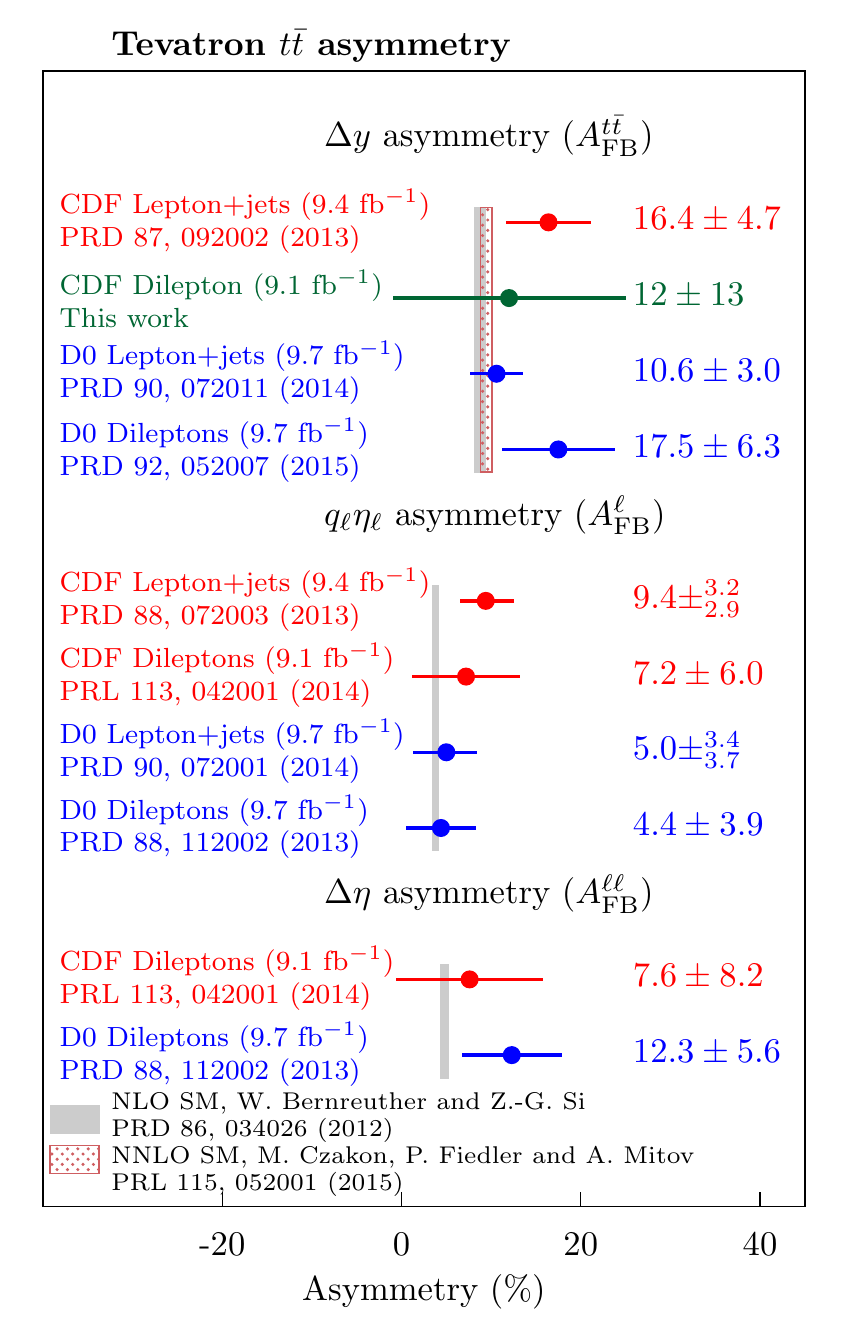}
    \caption{A comparison of all inclusive top-quark-pair forward--backward asymmetry results from the Tevatron with the NLO and NNLO SM predictions.}
    \label{fig:TevAFBComp}
\end{figure}

We finally determine the $\afbtt$ value by applying the parton-level extraction to data. 
Figure~\ref{fig:AFBResult} shows the posterior-probability density of the inclusive $\afbtt$. A Gaussian function is fitted to the core of the distribution to determine the central value of $\afbtt$ and its statistical uncertainty. Including the systematic uncertainties summarized in Table~\ref{tab:systematics}, the parton-level inclusive $\afbtt$ is measured to be
\begin{equation}
    \afbtt = 0.12 \pm 0.11 (\stat) \pm 0.07 (\syst) = 0.12\pm 0.13.
    \label{eq:AFBResult}
\end{equation}
The result is compared to previous $\afbtt$ measurements performed at the Tevatron and NLO and NNLO SM predictions in Fig.~\ref{fig:TevAFBComp}~\cite{Bernreuther:2012sx,Czakon:2014xsa}. 
No significant deviation is observed.

\begin{figure}[hbtp]
    \includegraphics[width=\columnwidthfigure]{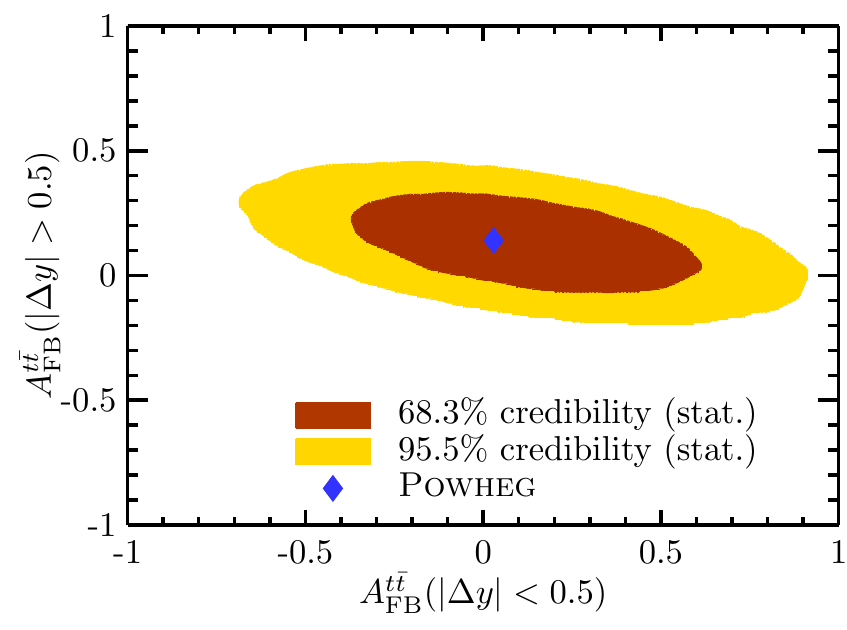}
    \caption{Two-dimensional posterior-probability-density distribution of $\afbtt (|\dy|>0.5)$ vs. $\afbtt (|\dy|<0.5)$. 
    }
    \label{fig:AFBdY2D}
\end{figure}

The posterior-probability densities of $\afbtt (|\dy|<0.5)$ and $\afbtt (|\dy|>0.5)$ are also Gaussian distributed. Gaussian functions are fitted to the core of the distributions to determine the central values of $\afbtt (|\dy|<0.5)$ and $\afbtt (|\dy|>0.5)$ and their statistical uncertainties. Including the systematic uncertainties summarized in Table~\ref{tab:systematics}, the parton-level values for $\afbtt$ vs. $|\dy|$ are measured to be
\begin{align}
    \begin{split}
        \afbtt(|\dy|<0.5) &= 0.12 \pm 0.33 (\stat) \pm 0.20 (\syst)\\
                          &= 0.12 \pm 0.39,
    \end{split}
    \\
    \begin{split}
        \afbtt(|\dy|>0.5) &= 0.13 \pm 0.13 (\stat) \pm 0.11 (\syst)\\
                          &= 0.13 \pm 0.17,
    \end{split}
    \label{eq:AFBdYFinal}
\end{align}
consistent with the predictions from \textsc{powheg} MC simulation of $0.017\pm0.001$ and $0.081\pm0.001$, respectively. The uncertainties on the predictions are due to the limited number of generated events in the MC simulation.
The uncertainty for $|\dy|<0.5$ is larger because of the large bin migrations in that region, which reduce the statistical power of the data. 
Figure~\ref{fig:AFBdY2D} shows the two-dimensional posterior-probability-density distribution of $\afbtt$ in the two $|\dy|$ regions, which shows that the two measurements are anticorrelated as expected. The correlation is $-0.44$.

\begin{figure}
    \includegraphics[width=\columnwidthfigure]{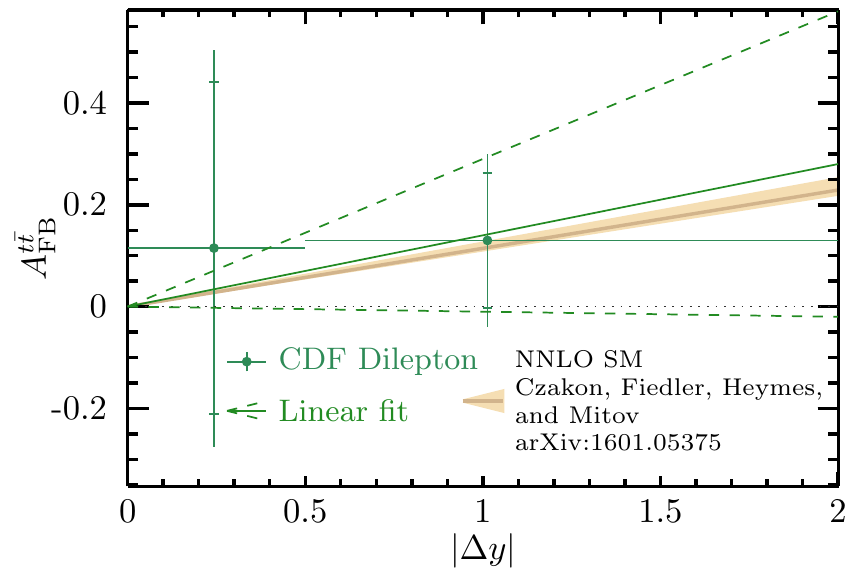}
    \caption{Comparison of the $\afbtt$ vs. $|\dy|$ dilepton results with the NNLO SM prediction~\cite{Czakon:2016ckf}. The data points are displayed at the bin centroids predicted by the \textsc{powheg} MC sample. The linear fit with zero intercept yields a slope of $0.14\pm0.15$.}
    \label{fig:AFBdYComp}
\end{figure}

To determine the slope of $\afbtt$ vs. $|\dy|$, we display the data points at the bin centroids predicted by the \textsc{powheg} MC sample and fit the two differential $\afbtt$ results with a linear function with zero intercept~\cite{Aaltonen:2012it}, taking all uncertainties with their correlations into account. The resultant slope is $\alpha=0.14\pm0.15$. Figure~\ref{fig:AFBdYComp} shows a comparison of the $\afbtt$-vs-$|\dy|$ results of this measurement with the NNLO SM prediction of $\alpha=0.114^{+0.006}_{-0.012}$~\cite{Czakon:2016ckf}. The result is consistent with the prediction.

\section{\label{sec:CDFComb}Combination of inclusive and differential $\afbtt$ results}
We combine the dilepton results with results obtained in the lepton+jets final state and reported in Ref.~\cite{Aaltonen:2012it}. The inclusive $\afbtt$ measured in the lepton+jets final state is $0.164\pm0.039(\stat)\pm0.026(\syst)$, with the slope of $\afbtt$ as a function of $|\dy|$ measured to be $0.253\pm0.062$. 

\begin{figure*}[htbp]
    \includegraphics[width=\columnwidthfigure]{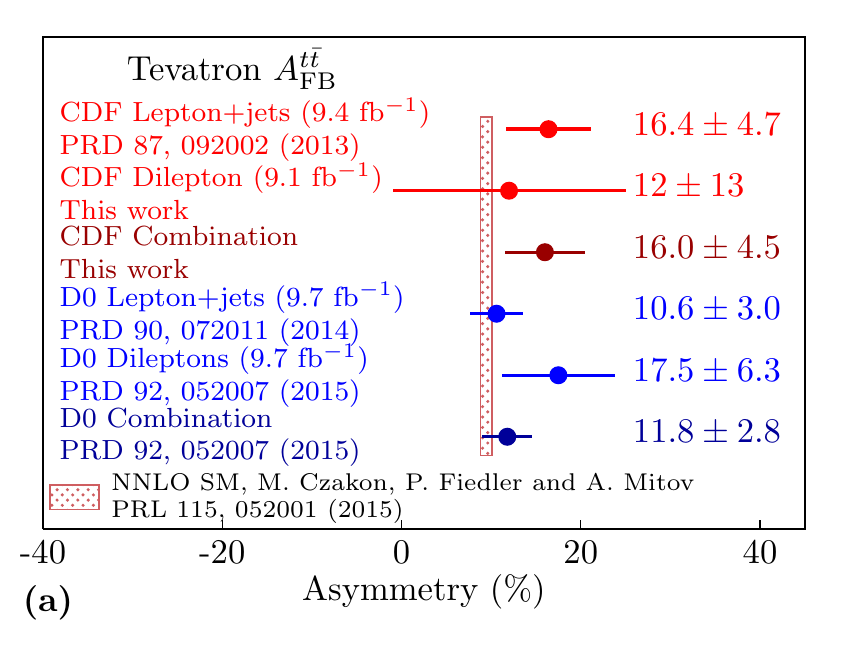}
    \includegraphics[width=\columnwidthfigure]{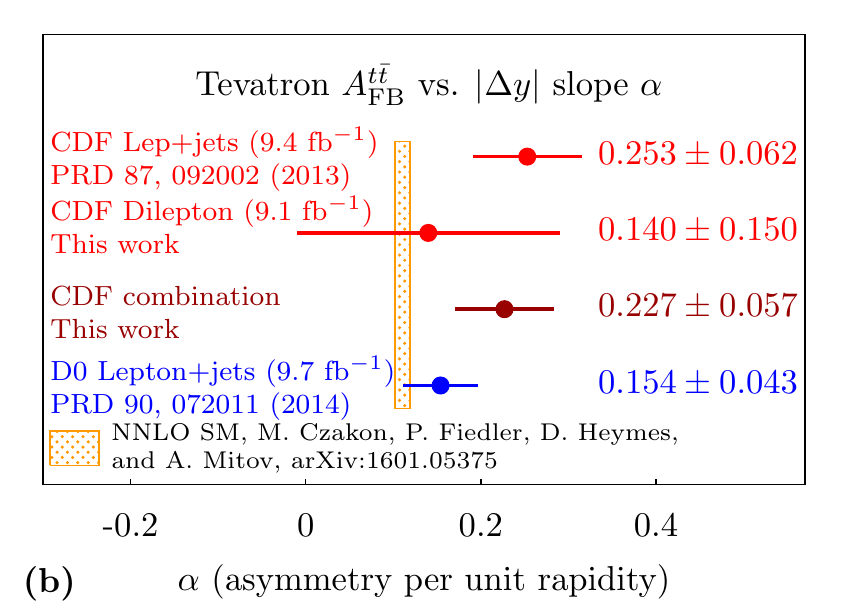}
    \caption{Comparison of the combined inclusive $\afbtt$ and the slope $\alpha$ of $\afbtt$ vs. $|\dy|$ with all other Tevatron measurements and the NNLO SM calculations.}
    \label{fig:AFBcomparison}
\end{figure*}

The treatment of the correlations of the statistical and systematic uncertainties between the two measurements follows Ref.~\cite{Aaltonen:2014eva}. 
Here, we summarize the various uncertainties and how they are combined. Since the two measurements are based on statistically independent samples, the statistical uncertainties are uncorrelated. While the two measurements share a small portion of the background source ($W$+jets), the background systematic uncertainties are mainly caused by the uncertainties in the shape of the background $\dy$ distributions, which are uncorrelated between the two measurements, and thus the associated uncertainties are treated as uncorrelated.
The correction and parton-level $\afbtt$ estimation procedures are different in the two measurements. Thus, the corresponding uncertainties are treated as uncorrelated.
The effects due to the uncertainties in the parton shower model, the jet-energy scale, the initial- and final-state radiation, the color-reconnection model, and the parton-distribution functions are estimated in identical ways. Thus, they are treated as fully correlated.
Table~\ref{tab:Correlation} summarizes the uncertainties and the correlations in both inclusive $\afbtt$ measurements. The combination of the inclusive $\afbtt$ is based on the best-linear-unbiased estimator~\cite{Lyons1988110,*PhysRevD.41.982,*Valassi2003391}.  With these uncertainties and the correlations, the combined value is
\begin{equation}
    \afbtt=0.160\pm0.045.
\end{equation}
The weights of the lepton+jets result and the dilepton result are 91\% and 9\%, respectively. The correlation between the two results is 10\%. The comparison of the combined result with other measurements and SM calculations is shown in Fig.~\ref{fig:AFBcomparison}(a).

\begin{table*}[hbtp]
    \caption{Table of uncertainties for the inclusive and differential $\afbtt$ measurements in the lepton+jets~\cite{Aaltonen:2012it} and the dilepton final states and their correlations.}
    \label{tab:Correlation}
    \TableUncertaintyCorrelation
\end{table*}

For the differential $\afbtt$, rather than combining the data, we perform a simultaneous fit for the slope $\alpha$ of the differential $\afbtt$ as a function of $|\dy|$ using both sets of data points (four in the lepton+jets final state and two in the dilepton final state). The position of the bin centroids expected by the \textsc{powheg}-MC sample and the $\afbtt$ in those bins are summarized in Table~\ref{tab:BinCentroidAFB} with the eigenvalues and the eigenvectors of the corresponding covariance matrix in Table~\ref{tab:CovMat}. The treatment of the correlations in the covariance matrix follows that used in the combination of the inclusive $\afbtt$, summarized in Table~\ref{tab:Correlation}.
The simultaneous fit is obtained by minimizing a $\chi^2$-like quantity defined as
\begin{widetext}
    \begin{equation}
        \chi^{2}=\sum_{i=1}^{6} \sum_{j=1}^{6}(\afbtt[i]-\alpha|\dy|[i])\mathrm{C}^{-1}[i][j](\afbtt[j]-\alpha|\dy|[j]),
    \end{equation}
\end{widetext}
where $|\dy|[i]$ and $\afbtt[i]$ are the $i$th bin centroids and the $\afbtt(|\dy|)$ values shown in Table~\ref{tab:BinCentroidAFB} respectively, $\mathrm{C}^{-1}[i][j]$ is the corresponding element of the inverse of the covariance matrix whose eigenvalues and eigenvectors are shown in Table~\ref{tab:CovMat}, and $\alpha$ is the slope determined by the fit. The result is $\alpha=0.227\pm0.057$, which is $2.0\sigma$ larger than the NNLO SM prediction of $0.114^{+0.006}_{-0.012}$~\cite{Czakon:2016ckf}. A comparison of the slope $\alpha$ with all results from CDF and D0 and the NNLO SM prediction is shown in Fig.~\ref{fig:AFBcomparison}(b).

\begin{table*}[hbtp]
    \caption{Bin centroids and the differential $\afbtt$ in the $\afbtt$ vs. $|\dy|$ measurements in both the lepton+jets and the dilepton final states.}
    \label{tab:BinCentroidAFB}
    \TableDifferentialCombinationBinCentroidAFB
\end{table*}

\begin{table*}[hbtp]
    \caption{The eigenvalues and eigenvectors of the covariance matrix for the $\afbtt$ vs. $|\dy|$ measurements in both the lepton+jets and the dilepton final states. Each row contains first an eigenvalue, then the error eigenvector that corresponds to that eigenvalue.}
    \label{tab:CovMat}
    \TableDifferentialCombinationCovMat
\end{table*}

\section{\label{sec:conclusion} Conclusion}

We measure parton-level the forward--backward asymmetries in the production of top quark and antiquark pairs reconstructed in the final state with two charged leptons using the full data set of $\sqrt{s} = 1.96$ TeV proton-antiproton collisions collected by the CDF II detector and corresponding to an integrated luminosity of $9.1~\fbm$. We measure the asymmetries inclusively and as functions of rapidity difference between top quark and antiquark. 
The results from the dilepton final state are $\afbtt = 0.12 \pm 0.13$, $\afbtt(|\dy|<0.5)=0.12\pm0.39$, and $\afbtt(|\dy|>0.5)=0.13\pm0.17$. A linear fit with zero intercept to the differential $\afbtt$ as a function of $|\dy|$ yields a slope of $\alpha=0.14\pm0.15$. 
We combine the above results with previous CDF results based on the final state with a single charged lepton and hadronic jets~\cite{Aaltonen:2012it}. 
The inclusive $\afbtt$ yields a value of $\afbtt=0.160\pm0.045$, which is consistent with the NNLO SM prediction of $0.095\pm0.007$~\cite{Czakon:2014xsa} within $1.5\sigma$. The simultaneous linear fit for $\afbtt$ as a function of $|\dy|$ with zero intercept yields a slope of $\alpha=0.227\pm0.057$, which is $2.0\sigma$ higher than the NNLO SM prediction~\cite{Czakon:2016ckf}. These are the final results of the CDF program for the exploration of top forward-backward asymmetries and, along with previous findings, show consistency with the predictions of the standard model at next-to-next-to-leading order.

\begin{acknowledgments}
    We thank the Fermilab staff and the technical staffs of the
    participating institutions for their vital contributions. This work
    was supported by the U.S. Department of Energy and National Science
    Foundation; the Italian Istituto Nazionale di Fisica Nucleare; the
    Ministry of Education, Culture, Sports, Science and Technology of
    Japan; the Natural Sciences and Engineering Research Council of
    Canada; the National Science Council of the Republic of China; the
    Swiss National Science Foundation; the A.P. Sloan Foundation; the
    Bundesministerium f\"ur Bildung und Forschung, Germany; the Korean
    World Class University Program, the National Research Foundation of
    Korea; the Science and Technology Facilities Council and the Royal
    Society, United Kingdom; the Russian Foundation for Basic Research;
    the Ministerio de Ciencia e Innovaci\'{o}n, and Programa
    Consolider-Ingenio 2010, Spain; the Slovak R\&D Agency; the Academy
    of Finland; the Australian Research Council (ARC); and the EU community
    Marie Curie Fellowship Contract No. 302103.
\end{acknowledgments}

\pagebreak

\bibliographystyle{apsrev4-1-JHEPfix}
\bibliography{citations}

\begin{thebibliography}{53}
\bibitem [{geo()}]{geometry}%
  \BibitemOpen
  \href@noop {} {}\bibinfo {note} {We use a cylindrical coordinate system with
  the origin at the center of the CDF II detector, $z$ pointing in the
  direction of the proton beam, $\theta$ and $\phi$ representing the polar and
  azimuthal angles, respectively, and pseudorapidity defined by $\eta = - \rm
  ln$ $\rm tan(\theta/2)$. Rapidity is defined as $y=\frac{1}{2}\ln
  \frac{E+p_{z}}{E-p_{z}}$, with positive $z$ defined along the proton
  direction.}\BibitemShut {Stop}%
\bibitem [{MET()}]{MET}%
  \BibitemOpen
  \href@noop {} {}\bibinfo {note} {The longitudinal momentum $p_{z}$ is defined
  to be $p\cos\theta$. The transverse momentum $p_T$ (transverse energy
  $E_{T}$) is defined to be $p\sin\theta$ ($E\sin\theta$). The missing
  transverse energy $\vec{\met}$ is defined to be
  $-\Sigma_{i}E_{T}^{i}$\^{n}$_{i}$ where $i$ is the calorimeter tower number
  with $|\eta|<3.6$, \^{n}$_{i}$ is a unit vector perpendicular to the beam
  axis and pointing at the $i$th calorimeter tower, and $\met$ is the magnitude
  of $\vec{\met}$}\BibitemShut {NoStop}%
\end{thebibliography}

\end{document}